\documentclass[journal]{IEEEtran}
\usepackage{lipsum,afterpage,refcount}
\usepackage{siunitx}
\usepackage[T1]{fontenc}
\usepackage{booktabs}  
\usepackage{graphicx}
\usepackage{fancyhdr}
\usepackage{setspace}
\usepackage{url}
\usepackage{hyperref}
\usepackage{amsthm}
\usepackage{bm}
\usepackage{mathtools}
\usepackage{flexisym}
\usepackage{multirow}
\usepackage{color}
\usepackage{tablefootnote}
\usepackage[flushleft]{threeparttable}
\usepackage[table]{xcolor} 
\usepackage{array}
\usepackage{stfloats}
\usepackage{changepage} 
\usepackage{algpseudocode}
\usepackage{amssymb}
\usepackage{enumitem}
\usepackage{verbatim}
\usepackage{amsmath}
\usepackage{algorithm}
\usepackage{tikz}
\usepackage{bbding}
\usepackage{xspace}
\usepackage{academicons}
\usepackage{textcomp}
\usepackage{xparse}

\newcommand*{\fun}{\NewDocumentCommand}

\makeatletter
\global\let\tikz@ensure@dollar@catcode=\relax
\makeatother

\DeclarePairedDelimiterX\set[1]\lbrace\rbrace{#1}

\fun \fg{m}{\Figure \ref{#1}\xspace}
\fun \hw{}{HW}
\fun \hwSpc{}{\hw\xspace}

\fun{\DSPnote}{}{\item[1] Determines if the design includes an optimization to
pack two smaller-bit multiplications onto the 18-bit multipliers of the DSPs.}
\fun{\tabThpt}{}{Throughput (GOPS)}
\fun{\kmmMacUt}{}{$\frac{\tx{\n 8-bit mults} / \mbit*\tx{multiplier}} {\tx{clock cycle}}$ \tnote{2}}
\fun{\kmmMacUtFFIP}{}{$\frac{\tx{\n 8-bit mults} / \mbit*\tx{multiplier}} {\tx{clock cycle}}$ \tnote{2}}
\fun {\kmmMacUtExpl}{}{\item[2] Multiplier compute efficiency, used to compare the amount of computational work being performed per compute area regardless of the input bitwidths or clock frequency, defined in \eq{kmm:mu} from \secn{kmm:sec:mu}, relevance explained in \secn{kmm:metrics}.}

\fun \winoConv{}{\cite{liu2021winocnn}, \cite{lavin2016fast}\xspace}
\fun \multDominant{}{\cite{liu2021winocnn}, \cite{jouppi2017datacenter}, \cite{norrie2021design}\xspace}
\fun \multComplexity{}{\cite{Lakshmi2022}, \cite{guo2019dl}, \cite{Pekmestzi1999}\xspace}
\fun \citeArea{}{\cite{moradi2009ultra}, \cite{kawai2014fully}, \cite{cai2019ultra}\xspace}

\fun \citePsML{}{\cite{li2022precision}, \cite{li2022low}, \cite{umuroglu2018bismo}\xspace}

\fun \Area{}{Area}
\fun \areaInt{mo}{\tx{\Area}({#1})}
\fun \areaB{mO{\ww}}{\tx{\Area}(\tx{#1}\bits{#2})}
\fun \area{mO{\AU}}{\areaInt{#1}[#2]}
\fun \areaT{mO{\AU}}{\areaInt{\tx{#1}}[#2]}
\fun \bits{mo}{\IfNoValueTF{#2}{^{[#1]}}{}}
\newcommand*{\algsize}{\small}
\fun \bld{}{\mathbf}
\fun \win{}{w}
\fun \wm{}{m}
\fun \ww{}{w}
\fun \wa{}{w_a}
\fun \waT{}{w_a}
\fun \wP{}{w_p}
\fun \wPT{}{w_p}
\fun \Hw{m}{\ceil{#1/2}}
\fun \Hwf{m}{\lfloor{#1/2}\rfloor}
\fun \hW{E{>}{{\dw}}}{\ceil{#1/2}}
\fun \dw{}{\win}
\fun \dm{}{\wm}
\fun \klog{}{^{\tx{log}_23}}

\fun \AU{}{AU}
\fun \autx{}{\AU\xspace}
\fun \au{}{\tx{ \AU}}

\fun \dd{}{d}
\fun \nn{}{n}
\fun \one{}{1}
\fun \dFpSub{}{1}
\fun \dPsSub{}{2}

\fun \eqFpSup{O{\dw}O{\dm}o}{\bits{#1}[#3]}
\fun \eqPsSup{O{\dw}O{\dm}o}{\bits{#1}[#3]}

\fun \kffip{}{FFIP+\KMM}

\fun \KMM{}{KMM}
\fun \MM{}{MM}
\fun \MMN{}{MM}
\fun \SM{}{SM}
\fun \KSM{}{KSM}
\fun \ksMM{}{KSMM}
\fun \ksmmOne{}{}

\fun \fpW{O{\win}}{#1}
\fun \psW{O{\win}}{#1}

\fun \psPrefix{}{}
\fun \fpPrefix{}{}

\fun \eqTxtAlg{ooooooooo} {\IfBooleanTF {#8}
  {\IfBooleanTF {#5}
    {\tx{{#1}}_{#2}}
    {{#1}$_{#2}$\xspace}}
  {\IfBooleanTF {#5}
    {\tx{#9{#1}}_{#2}#6[#3][#4][#7]}
    {#9{#1}$_{#2}#6[#3][#4][#7]$\xspace}}}

\fun \PsFp{oooooooo} {\IfBooleanTF {#1}
  {\eqTxtAlg[#2][#3][#4][#5][#6][\eqFpSup][#7][#8][\fpPrefix]}
  {\eqTxtAlg[#2][#3][#4][#5][#6][\eqPsSup][#7][#8][\psPrefix]}}

\fun \KMMthpt{E{/}{{\ww}}s}{\IfBooleanTF{#2}{\alg.\MM'*/#1?\one<{} \tx{ throughput}}{\mm throughput\xspace}}
\fun \mbit{s}{}
\fun \Ndigit{}{n-Digit\xspace}
\fun \ndigit{}{n-digit\xspace}
\fun \nx{}{}
\fun \nxT{}{}
\fun \hex{s}{\IfBooleanTF{#1}{\tx{0x}}{0x}}

\fun \mm{E{/?}{{\ww}{\one}}s}{\IfBooleanTF{#3}
{\alg.\MM'*/{#1}?{#2}<{}}{\alg.\MM'/{#1}?{#2}<{}}}

\fun \mmn{E{/?}{{\ww}{\nx}}s}{\IfBooleanTF{#3}
{\alg.\MMN'*/{#1}?{#2}<{}}{\alg.\MMN'/{#1}?{#2}<{}}}

\fun \sm{E{/?}{{\ww}{\nx}}s}{\IfBooleanTF{#3}
{\alg.\SM'*/{#1}?{#2}<{}}{\alg.\SM'/{#1}?{#2}<{}}}

\fun \kmm{E{/?}{{\ww}{\nx}}s}{\IfBooleanTF{#3}
{\alg.\KMM'*/{#1}?{#2}<{}}{\alg.\KMM'/{#1}?{#2}<{}}}

\fun \ksmm{E{/?}{{\ww}{\nx}}s}{\IfBooleanTF{#3}
{\alg.\ksMM*'/{#1}?{#2}\<{}}{\alg.\ksMM'/{#1}?{#2}<{}}}

\fun \ksm{E{/?}{{\ww}{\nx}}s}{\IfBooleanTF{#3}
{\alg.\KSM'*/{#1}?{#2}<{}}{\alg.\KSM'/{#1}?{#2}<{}}}

\fun \kmmArch{E{/?}{{\win}{\nx}}}{\alg/{#1}?{#2}<{}}
\fun \fpMm{E{/?}{{\win}{\one}}}{\alg.\MM'/{#1}?{#2}<{}}
\fun \fpKmm{E{/?}{{\win}{\nx}}}{\alg.\KMM'/{#1}?{#2}<{}}
\fun \fpKsmm{E{/?}{{\win}{\nx}}}{\alg.\ksMM'/{#1}?{#2}<{}}
\fun \psKsmm{E{/?}{{\win}{\nxT}}}{\alg.\ksMM'/{#1}?{#2}<{}}
\fun \psKmm{E{/?}{{\win}{\nxT}}}{\alg/{#1}?{#2}<{}}
\fun \psMm{E{/?}{{\win}{\nxT}}}{\alg.\MMN/{#1}?{#2}<{}}

\fun \quadratically{}{quadratically\xspace}

\fun \alg{E{.}{{\KMM}}t'sE{?>/<}{{\dPsSub}{\dm}}t!}
     {\PsFp[#2][#1][#4][\dW[#6][#2]][#5][#3][#7][#8]}

\fun \dW{oo} {\IfBooleanTF {#2}
    {\fpW[#1]}
    {\psW[#1]}
}

\fun \dWW{E{4}{}}{
  \IfNoValueTF {#1}
               {\fpW[0]}
               {\fpW[1]}
}

\fun\caredd{mO{}}{\frac {\mathcal{O}_{#2}( \tx{baseline alg} )} {\mathcal{O}_{#2}{( \tx{\hwSpc alg} )}}}

\newcommand*{\cred}[2]{\mathcal{O}( \tx{baseline alg} ) / \mathcal{O}{( \tx{\hwSpc alg} )}}

\newcommand*{\n}{\xspace}

\newcommand*{\KMMop}[2]{#1\bits{#2}}
\newcommand*{\mult}[1]{\KMMop{\tx{MULT}}{#1}}
\newcommand*{\add}[1]{\KMMop{\tx{ADD}}{#1}}
\newcommand*{\accum}[1]{\KMMop{\tx{ACCUM}}{#1}}
\newcommand*{\shift}[1]{\KMMop{\tx{SHIFT}}{#1}}
\newcommand*{\ff}[1]{\KMMop{\tx{FF}}{#1}}
\fun \eqff{}{\tx{FF}}
\newcommand*{\ccc}[2]{C_{#2}(#1)}
\fun{\cc}{mO{}}{\ccc{#1}{#2}}
\fun{\opcc}{mO{}}{#1}

\newcommand{\orcid}[1]{\href{kmm:https://orcid.org/#1}{\textcolor[HTML]{A6CE39}{\includegraphics[width=1.7ex]{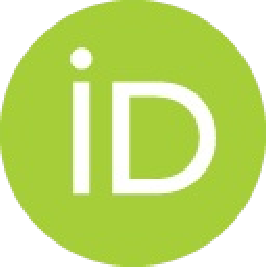}}}}

\newcommand{\eq}[1]{(\ref{#1})\xspace}
\newcommand{\eqs}{}
\newcommand*{\Figure}{Fig.\xspace}
\newcommand*{\fig}[1]{\Figure \ref{#1}}

\fun{\secn}{m}{Section \ref{#1}\xspace}
\fun{\Secn}{m}{Section \ref{#1}\xspace}
\newcommand*{\sections}{Sections\xspace}

\newcommand*{\eALMs}{118K}
\newcommand*{\eRegs}{311K}
\newcommand*{\eMems}{1782}
\newcommand*{\eDSPs}{1072}
\newcommand*{\eFreq}{388}
\newcommand*{\eResNetAGOPS}{2529}
\newcommand*{\eResNetBGOPS}{2752}
\newcommand*{\eResNetCGOPS}{2838}

\newcommand*{\lba}{\left(}
\newcommand*{\rba}{\right)}
\newcommand*{\lbb}{\left(}
\newcommand*{\rbb}{\right)}
\newcommand*{\lb}{(}
\newcommand*{\rb}{)}

\newcommand*{\ea}{et al.\xspace}

\newcommand*{\gx}{Arria 10 GX 1150\xspace}
\newcommand*{\sx}{Arria 10 SX 660\xspace}

\newcommand*{\tx}[1]{\text{#1}}
\newcommand*{\spc}{\,}
\newcommand*{\spcc}{\,\,}

\newcommand*{\x}{$\times$\xspace}
\newcommand*{\by}{$\times$}

\DeclarePairedDelimiter{\ceil}{\lceil}{\rceil}
\AtBeginDocument{
  \catcode`_=12
  \begingroup\lccode`~=`_
  \lowercase{\endgroup\let~}\sb
  \mathcode`_="8000
}
\newcolumntype{L}{>{\centering\arraybackslash}p{0.58cm}}
\newcolumntype{Z}{>{\centering\arraybackslash}p{.95cm}}
\newcolumntype{A}{>{\centering\arraybackslash}p{.7cm}}
\newcolumntype{V}{>{\centering\arraybackslash}p{.7cm}}
\newcolumntype{B}{>{\centering\arraybackslash}p{.6cm}}
\newcolumntype{C}{>{\centering\arraybackslash}p{.85cm}}
\newcolumntype{D}{>{\centering\arraybackslash}p{1.3cm}}
\newcolumntype{H}{>{\setbox0=\hbox\bgroup}c<{\egroup}@{}}

\hypersetup{
  colorlinks=true,
  linkcolor=blue!50!blue,
  citecolor=blue!50!blue,
  urlcolor=black!70!black
}

\newcommand\copyrighttext{%
  \scriptsize \textcopyright
  2025 IEEE. Personal use of this material is permitted. Permission
  from IEEE must be obtained for all other uses, in any current or future
  media, including reprinting/republishing this material for advertising or
  promotional purposes, creating new collective works, for resale or
  redistribution to servers or lists, or reuse of any copyrighted
  component of this work in other works.
  Accepted for publication in IEEE Transactions on Computers. DOI: 10.1109/TC.2025.3525606}
\newcommand\copyrightnotice{%
  \begin{tikzpicture}[remember picture,overlay]
    \node[anchor=south,yshift=0pt] at (current page.south) {\fbox{\parbox{\dimexpr\textwidth-\fboxsep-\fboxrule\relax}{\copyrighttext}}};
  \end{tikzpicture}
  \vspace{-11.94pt}
}

\begin{document}

\title{Karatsuba Matrix Multiplication and its Efficient Custom Hardware Implementations}

\author{Trevor~E.~Pogue~\orcid{0000-0002-6791-3758} and Nicola~Nicolici~\orcid{0000-0001-6345-5908},~\IEEEmembership{Senior Member,~IEEE}
  \IEEEcompsocitemizethanks{\IEEEcompsocthanksitem T. E. Pogue and N. Nicolici are with the Department of Electrical and Computer Engineering, McMaster University, Hamilton, ON, L8S 4L8, Canada \protect\\
    Email: poguete@mcmaster.ca; nicolici@mcmaster.ca
}}

\IEEEpeerreviewmaketitle

\IEEEtitleabstractindextext{%
\begin{abstract}
    While the Karatsuba algorithm reduces the complexity of large integer
multiplication,
the extra additions required minimize its benefits
for smaller integers of more commonly-used bitwidths.
In this work, we propose the extension of the scalar Karatsuba multiplication
algorithm to matrix multiplication, showing how this maintains the reduction in
multiplication complexity of the original Karatsuba algorithm while reducing the
complexity of the extra additions.
Furthermore, we propose new matrix multiplication hardware architectures for
efficiently exploiting this extension of the Karatsuba algorithm in custom
hardware.
We show that the proposed algorithm and hardware architectures can provide real
area or execution time improvements for integer matrix multiplication compared
to scalar Karatsuba or conventional matrix multiplication algorithms, while also
supporting implementation through proven systolic array and conventional
multiplier architectures at the core.
We provide a complexity analysis of the algorithm and architectures
and evaluate the proposed designs both in isolation
and in an end-to-end deep learning accelerator system compared to baseline designs and prior
state-of-the-art works implemented on the same type of compute platform,
demonstrating their ability to increase the performance-per-area of matrix
multiplication hardware.

\end{abstract}
}

\maketitle
\IEEEdisplaynontitleabstractindextext
\IEEEpeerreviewmaketitle
\section{Introduction}
\copyrightnotice
\IEEEPARstart{T}{he} demand for optimized hardware acceleration of general
matrix multiplication (GEMM) continues to drive innovation in the field of
hardware design for exploiting the inherent parallelism to speed up computation.
At a certain point, however, after the known parallelism and system-level
optimizations are exhausted and technology scaling slows to a halt, there is an
accelerator wall which limits further progress on the implementation side
\cite{fuchs2019accelator}.
A less-explored direction for continuing advancement beyond this wall is through
reducing the workload at the algebraic level, by computing the same result from
a re-arranged compute pattern requiring fewer or cheaper operations to be
performed in hardware.

Multiply-accumulate (MAC) units are commonly the area-dominant computational
resource in GEMM and deep learning accelerators \multDominant, and due to this,
an accelerator's throughput can be directly limited by how many multipliers its
hardware budget can afford. As a result, surpassing this performance per
multiplier limit has been focused on recently with minimal filtering algorithms
applied to convolutional neural networks \winoConv, as well fast inner-product
algorithms for GEMM and machine learning workloads \cite{pogue2024fast}.
Along this same direction, the Karatsuba algorithm \cite{karatsuba1962} can also
theoretically be used to reduce the complexity of integer multiplication.
However, the extra addition operations it introduces
can increase its execution speed in general-purpose computers or limit its area
reduction in
custom multiplier circuits for smaller integers of more commonly-used bitwidths
\cite{jain2021approxkarat}, \cite{jain2021boothkarat}.

In this work, we show how the scalar Karatsuba multiplication algorithm can be
extended to integer matrix multiplication, after which the impact and complexity
of the extra additions is reduced.
Furthermore, we investigate and present new fixed-precision and
precision-scalable hardware architectures for efficiently exploiting the
Karatsuba algorithm extended to matrix multiplication (referred to as Karatsuba
matrix multiplication or \kmm), showing how the proposed algorithm and hardware
architectures can provide real
area or execution time reductions for integer matrix multiplication compared to
scalar Karatsuba or conventional matrix multiplication.

The proposed architectures can also be implemented using proven systolic array
and conventional multiplier architectures at their core, maintaining all the
implementation benefits of these architectures.
Systolic arrays, which we will also refer to as matrix multiplication units
(MXU)s for convenience, are an effective choice for use in GEMM accelerators as
they significantly reduce the required memory traffic and can reach high clock
frequencies due to their short and regular interconnects. Systolic-array
architectures have been used in state-of-the-art
GEMM and deep learning accelerators such as the Tensor Processing Unit (TPU)
\cite{jouppi2017datacenter}, \cite{norrie2021design}, \cite{jouppi2023tpu},
among others \cite{pogue2024fast}, \cite{zhang2019caffeine}.

In summary, our key contributions are the following:
\begin{itemize}
\item We propose the Karatsuba matrix multiplication (\kmm) algorithm and carry
  out a complexity analysis of the algorithm compared to conventional scalar
  Karatsuba and matrix multiplication algorithms to facilitate further future
  investigations of potential applications and hardware implementations of \kmm.
  We also identify complexity shortcomings of \kmm that restrict its benefits in
  hardware and show how this is mitigated when \kmm is combined with an
  alternative accumulation algorithm.
\item We present a new family of hardware architectures for efficiently
  exploiting \kmm in custom hardware. We then model the area or execution time
  benefits of the \kmm architectures and evaluate the proposed architectures
  both in
  isolation and in an end-to-end accelerator system compared to baseline
  designs and prior state-of-the-art works implemented on the same type of
  compute platform.
\end{itemize}

\section{Background and Related Work}

\subsection{Notation}
We use the following notation throughput this article:
\begin{itemize}
\item ALG$_\nn\bits{\ww}$: An algorithm that operates on $\ww$-bit
  scalars or matrices with $\ww$-bit elements, where each scalar or matrix
  element is divided into $\nn$ digits. For example, \alg.\SM?2/8 represents a
  scalar multiplication (\sm) algorithm for operating on 8-bit 2-digit numbers
  where each digit is 4 bits wide, such as the multiplication between the
  hexadecimal values $\hex*12\times\hex*10 = \hex*120$.
  \begin{itemize}
  \item ALG$_\nn$ or ALG: The algorithm acronym may also be specified without the
    subscript $_\nn$ and/or superscript $\bits{\ww}$ when the number of digits
    and/or  input bitwidths are not directly relevant for the current
    context, and it may refer to the use of the algorithm for any value of $\nn$
    or $\ww$ for each missing subscript and/or superscript.
\end{itemize}
\item OPERATION$\bits{\ww}$: An arithmetic operation that works with $\ww$-bit
  values. For example, $\mult{\ww}$, $\add{\ww}$, $\accum{\ww}$
  represent a multiplication, addition, and accumulation of $\ww$-bit values,
  respectively, and $\shift{\ww}$ represents a left or right shift by $\ww$ bits.
\item x$\bits{a:b}$: The value contained in bits $a$ down to $b$ of a scalar
  $x$. For example, the value of bits 7 down to 4 in the hexadecimal number $\hex*AE$ is equal to $\hex*A$ and is written as $\hex*AE\bits{7:4} = \hex*A$. Similarly, $\hex*AE\bits{3:0} = \hex*E$.
\item $\cc{\tx{ALG}_\nn\bits{\ww}}$: The complexity of algorithm ALG in number of
  $\ww$-bit multiplications, additions, accumulations, and shift operations.
\item $\cc{\tx{ALG}_\nn}$: The complexity of algorithm ALG in number of arithmetic
  operations.
\item $r$: The number of recursion levels implemented in \ksm or \kmm, equal to $\ceil{\tx{log}_2n}$.
\item $\dd$: The height and width of two matrices being multiplied.
\end{itemize}

\subsection{Conventional \Ndigit Scalar Multiplication (\sm)}

\begin{figure}[]
  \centering
  \includegraphics[scale=1.1]{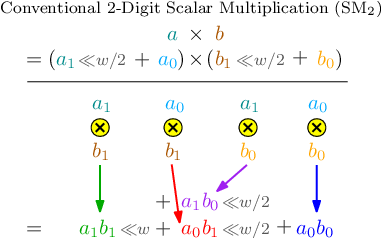}
  \caption{\sm?2 algorithm illustration.}
  \label{kmm:fig:sm}
\end{figure}

\begin{algorithm}[]
  \algsize
  \caption{Conventional \Ndigit Scalar Multiplication.}
  \label{kmm:alg:smn}
  \begin{algorithmic}[1]
    \Function{\alg.\SM'?\nn/\ww}{$a$, $b$}
    \If {($n > 1$)}
    \State $a_1 = a\bits{{\ww}{-}{1}{:}{\Hw{\ww}}}$
    \State $a_0 = a\bits{{\Hw{\ww}}{-}{1}{:}{0}}$
    \State $b_1 = b\bits{{\ww}{-}{1}{:}{\Hw{\ww}}}$
    \State $b_0 = b\bits{{\Hw{\ww}}{-}{1}{:}{0}}$
    \State $c_{1} = \alg.\SM'*?{n/2}/{\Hwf{\ww}}\lb a_{1},b_{1} \rb$
    \State $c_{10} = \alg.\SM'*?{n/2}/{\Hw{\ww}}\lb a_{1},b_{0} \rb$
    \State $c_{01} = \alg.\SM'*?{n/2}/{\Hw{\ww}}\lb a_{0},b_{1} \rb$
    \State $c_{0} = \alg.\SM'*?{n/2}/{\Hw{\ww}}\lb a_{0},b_{0} \rb$
    \State $c = c_{1} \ll \ww$
    \State $c \mathrel{+}= \lb c_{01} + c_{10} \rb \ll \Hw{\ww}$
    \State $c \mathrel{+}= c_{0}$
    \Else
    \State $c = a\times b$
    \EndIf
    \State return $c$
    \EndFunction
  \end{algorithmic}
\end{algorithm}

\fig{kmm:fig:sm} shows the conventional method for performing 2-digit scalar
multiplication where a $\ww$-bit multiplication is split into four smaller-bit
scalar multiplications before being summed to form the final product.
Algorithm \ref{kmm:alg:smn} shows the generalization of this, where n-digit
multiplication is performed by carrying out the same steps recursively for each
smaller-bit multiplication.

\subsection{Karatsuba Scalar Multiplication (\ksm)}

\begin{figure}[]
  \centering
  \includegraphics[scale=1.1]{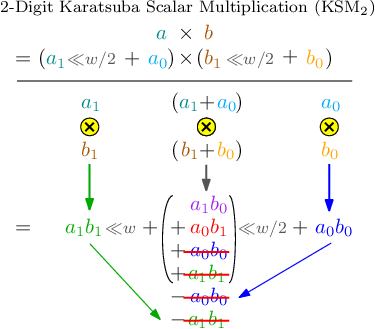}
  \caption{\ksm?2 algorithm illustration. Compared to \sm?2, \ksm?2 requires
    only 3 single-digit multiplications, however, it requires 3 more additions,
    increasing the overall operation count.}
  \label{kmm:fig:ksm}
\end{figure}

\fig{kmm:fig:ksm} shows the Karatsuba algorithm \cite{karatsuba1962} for 2-digit
scalar multiplication where a $\ww$-bit multiplication is split this time into
\textit{three} smaller-bit multiplications before being summed to form
the final product. Algorithm \ref{kmm:alg:ksm} shows the generalization of this,
where n-digit multiplication is performed by carrying out the same steps
recursively for each smaller-bit multiplication.

\begin{algorithm}[]
  \algsize
  \caption{\Ndigit Karatsuba Scalar Multiplication.}
 \label{kmm:alg:ksm}
  \begin{algorithmic}[1]
    \Function{\alg.\KSM'?\nn/\ww}{$a$, $b$}
    \If {($n > 1$)}
    \State $a_{1} = a\bits{\ww{-}{1}{:}\Hw{\ww}}$
    \State $a_{0} = a\bits{\Hw{\ww}{-}{1}{:}0}$
    \State $b_{1} = b\bits{\ww{-}{1}{:}\Hw{\ww}}$
    \State $b_{0} = b\bits{\Hw{\ww}{-}{1}{:}0}$
    \State $a_{s} = a_{1}+a_{0}$ \label{kmm:line:ksm-add-a}
    \State $b_{s} = b_{1}+b_{0}$ \label{kmm:line:ksm-add-b}
    \State $c_{1} = \alg.\KSM'*?{n/2}/{\Hwf{\ww}}\lb a_{1},b_{1} \rb$
    \State $c_{s} = \alg.\KSM'*?{n/2}/{\Hw{\ww}{+}{1}}\lb a_{s},b_{s} \rb$
    \State $c_{0} = \alg.\KSM'*?{n/2}/{\Hw{\ww}}\lb a_{0},b_{0} \rb$
    \State $c = c_{1} \ll \ww$
    \State $c \mathrel{+}= \lb c_{s} - c_{1} - c_{0} \rb \ll \Hw{\ww}$ \label{kmm:line:ksm-add-c1}
    \State $c \mathrel{+}= c_{0}$\label{kmm:line:ksm-add-c2}
    \Else
    \State $c = a\times b$
    \EndIf
    \State return $c$
    \EndFunction
  \end{algorithmic}
\end{algorithm}
\ksm-based low-bitwidth accurate integer multiplier circuits in prior works have
shown some area benefits for input bitwidths in the range of 64 bits or less,
with minimal area improvements in the smallest ranges of 16 bits
\cite{jain2021approxkarat}, \cite{jain2021boothkarat}.

\subsection{Conventional \Ndigit Matrix Multiplication (\mmn)}
A conventional matrix multiplication algorithm
computes $\bld{C} = \bld{A} \bld{B}$ for $\bld{A}$ of size $M \times K$ and $\bld{B}$
of size $K \times N$, where each element $c_{i,j}$ of $\bld{C}$ is calculated as
follows:
\begin{align} \label{eq:MM}
  c_{i,j} = \sum_{k=1}^{K} a_{i,k} b_{k,j} \,.
\end{align}

\begin{figure}[]
  \centering
  \includegraphics[scale=1.1]{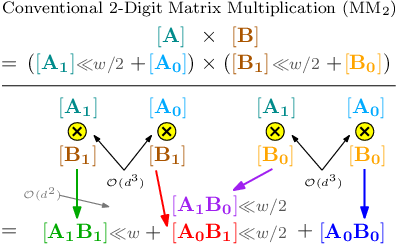}
  \caption{\mmn?2 algorithm illustration. The 4 single-digit matrix
    multiplications of complexity $\mathcal{O}\lb d^3\rb$ dominate the $\mathcal{O}\lb d^2\rb$
    complexity of the matrix additions.}
  \label{kmm:fig:mmn}
\end{figure}
\begin{algorithm}[]
  \algsize
  \caption{Conventional \Ndigit Matrix Multiplication.}
 \label{kmm:alg:mmn}
  \begin{algorithmic}[1]
    \Function{\alg.\MM'?\nn/\ww}{$\bld{A}$, $\bld{B}$}
    \If {($n > 1$)}
    \vspace*{0.1cm}
    \State $\bld{A_1} = \begin{bmatrix} a_{1,1}\bits{{\ww}{-}{1}{:}{\Hw{\ww}}}, & ... & a_{1,K}\bits{{\ww}{-}{1}{:}{\Hw{\ww}}} \\ ... & ...  & ... \\ a_{M,1}\bits{{\ww}{-}{1}{:}{\Hw{\ww}}}, & ... & a_{M,K}\bits{{\ww}{-}{1}{:}{\Hw{\ww}}} \end{bmatrix}$
    \State $\bld{A_0} = \begin{bmatrix} a_{1,1}\bits{{\Hw{\ww}}{-}{1}{:}{0}}, & ... & a_{1,K}\bits{{\Hw{\ww}}{-}{1}{:}{0}} \\ ... & ... & ... \\ a_{M,1}\bits{{\Hw{\ww}}{-}{1}{:}{0}}, & ... & a_{M,K}\bits{{\Hw{\ww}}{-}{1}{:}{0}} \end{bmatrix}$
    \State $\bld{B_1} = \begin{bmatrix} b_{1,1}\bits{{\ww}{-}{1}{:}{\Hw{\ww}}}, & ... & b_{1,N}\bits{{\ww}{-}{1}{:}{\Hw{\ww}}} \\ ... & ... & ... \\ b_{K,1}\bits{{\ww}{-}{1}{:}{\Hw{\ww}}}, & ... & b_{K,N}\bits{{\ww}{-}{1}{:}{\Hw{\ww}}} \end{bmatrix}$
    \State $\bld{B_0} = \begin{bmatrix} b_{1,1}\bits{{\Hw{\ww}}{-}{1}{:}{0}}, & ... & b_{1,N}\bits{{\Hw{\ww}}{-}{1}{:}{0}} \\ ... & ... & ... \\ b_{K,1}\bits{{\Hw{\ww}}{-}{1}{:}{0}}, & ... & b_{K,N}\bits{{\Hw{\ww}}{-}{1}{:}{0}} \end{bmatrix}$
    \vspace*{0.15cm}
    \State $\bld{C_{1}} = \alg.\MMN'*?{n/2}/{\Hwf{\ww}}\lb \bld{A_1},\bld{B_1} \rb$
    \State $\bld{C_{10}} = \alg.\MMN'*?{n/2}/{\Hw{\ww}}\lb \bld{A_1},\bld{B_0} \rb$
    \State $\bld{C_{01}} = \alg.\MMN'*?{n/2}/{\Hw{\ww}}\lb \bld{A_0},\bld{B_1} \rb$
    \State $\bld{C_{0}} = \alg.\MMN'*?{n/2}/{\Hw{\ww}}\lb \bld{A_0},\bld{B_0} \rb$
    \State $\bld{C} = \bld{C_{1}} \ll \ww$ \label{kmm:line:mmn-output-start}
    \State $\bld{C} \mathrel{+}= \lb \bld{C_{10}} + \bld{C_{01}} \rb \ll \Hw{\ww}$ \label{kmm:line:mmn-add1}
    \State $\bld{C} \mathrel{+}= \bld{C}_{0}$ \label{kmm:line:mmn-add2} \label{kmm:line:mmn-output-end}
    \Else
    \State $\bld{C} = \alg.\MM'*?\one/\ww \lb \bld{A},\bld{B} \rb$ \label{kmm:line:mm}
    \EndIf
    \State return $\bld{C}$
    \EndFunction
  \end{algorithmic}
\end{algorithm}

The method in \fig{kmm:fig:sm} can also be extended to matrix multiplication as
illustrated in \fig{kmm:fig:mmn}, where four separate partial-product matrix
multiplications are performed between matrices each containing bit slices of
every element, and they are later summed together to form the final matrix
product.
Algorithm \ref{kmm:alg:mmn} shows the generalization of this, where n-digit matrix
multiplication is performed by carrying out the same steps recursively for each
smaller-bit matrix multiplication.
The elements in matrices $\bld{A_0}$ and $\bld{B_0}$ contain the lower bits
(bits $\hW>\ww-1$ down to 0) of every element in the $\bld{A}$ and $\bld{B}$
matrices, while $\bld{A_1}$ and $\bld{B_1}$ contain the upper bits (bits $\ww-1$
down to $\hW>\ww$) of every element in matrices $\bld{A}$ and $\bld{B}$.
This allows for $\win$-bit matrix multiplication using smaller $\wm$-bit
multipliers. The \mm algorithm on line \ref{kmm:line:mm} of Algorithm
\ref{kmm:alg:mmn} is a conventional matrix multiplication algorithm such as
\eq{eq:MM}.

\subsection{Precision-Scalable Architectures}
Precision-scalable architectures allow a way to efficiently execute workloads
across multiple input precisions for applications where the input bitwidths are
expected to vary. Machine learning (ML) acceleration is one example of a
use-case for precision-scalable hardware architectures, where neural networks
can perform the majority of the inference on reduced-bitwidth operations with
little to no loss in accuracy but the bitwidths required to provide sufficient
accuracy vary across different deep neural network models, applications, and
between individual layers within the same neural network model
\cite{li2022precision}.
For example, some neural network models can be executed with high accuracy even
when performing the majority of the operations on small bitwidths, however, a
smaller portion of the layers still need to be computed on larger bitwidths to
preserve accuracy \cite{li2022precision}. Therefore, a fixed-bitwidth
accelerator must make a trade-off between either supporting only lower bitwidths
while reducing the model's accuracy, or supporting larger bitwidths for higher
accuracy but under-utilizing the MAC units during majority of computation as
most stages require only lower-bit inputs.

Precision-scalable architectures address this trade-off by providing
architectures which can more efficiently support execution of varying input
bitwidths \citePsML. One approach is to use MAC units consisting of multiple
smaller-bitwidth multipliers \cite{li2022low} which can either be individually
used to multiply/accumulate multiple smaller-bitwidth products, or they can be
reconfigured to collectively multiply/accumulate fewer larger-bitwidth products
per clock cycle.
Another type of approach referred to as bit-serial architectures
\cite{umuroglu2018bismo}, is to have MAC arrays which repeatedly perform
fixed-size smaller-bitwidth vector operations on different bit slices of the
vectors, summing up the separate vector products to get the final full-bitwidth
vector result.

The hardware algorithms used in prior works on precision-scalable architectures
\citePsML use variations of the \sm and \mmn algorithms shown in Algorithms
\ref{kmm:alg:smn} and \ref{kmm:alg:mmn} to combine partial products and compute
variable-bitwidth $\win$-bit matrix products using smaller $\wm$-bit
multipliers, where the number of $\wm$-bit multiplications and minimum possible
execution time if fully utilizing the $\wm$-bit multipliers scales quadratically
with the input bitwidths $\win$.
However, as shown later, the minimum possible execution time of a
precision-scalable \kmm architecture scales less than quadratically with the
input bitwidths $\win$.

\section{Karatsuba Matrix Multiplication (\kmm)}
In this section, we formally define \kmm, analyze its complexity compared to
conventional scalar Karatsuba and matrix multiplication algorithms, identify
complexity shortcomings of the \kmm algorithm that restrict its benefits in
hardware, and show how this is mitigated when combining \kmm with an alternative
accumulation algorithm.

\subsection{\kmm Definition}
\begin{figure}[]
  \centering
  \includegraphics[scale=1.1]{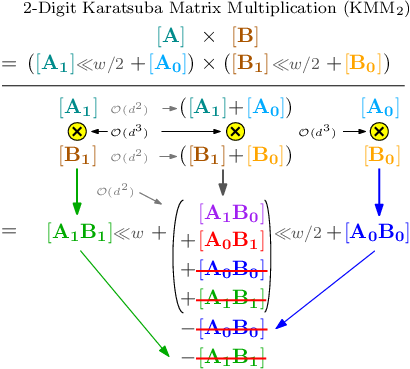}
  \caption{\kmm?2 algorithm illustration. Compared to the scalar algorithms
    \ksm?2 versus \sm?2, the increase in number of additions with complexity
    $\mathcal{O}\lb d^2\rb$ in \kmm?2 versus \mmn?2 is now
    insignificant relative to the reduction of 3 instead of 4 single-digit
    matrix multiplications of complexity $\mathcal{O}\lb d^3\rb$, allowing the overall
    \#operations in \kmm?2 to be less than conventional \mmn?2.
  }
  \label{kmm:fig:kmm}
\end{figure}
\begin{algorithm}[]
  \algsize
  \caption{\Ndigit Karatsuba Matrix Multiplication.}
 \label{kmm:alg:kmm}
  \begin{algorithmic}[1]
    \Function{\alg.\KMM'?\nn/\ww}{$\bld{A}$, $\bld{B}$}
    \If {($n > 1$)}
    \vspace*{0.1cm}
    \State $\bld{A_1} = \begin{bmatrix} a_{1,1}\bits{{\ww}{-}{1}{:}{\Hw{\ww}}}, & ... & a_{1,K}\bits{{\ww}{-}{1}{:}{\Hw{\ww}}} \\ ... & ... & ... \\ a_{M,1}\bits{{\ww}{-}{1}{:}{\Hw{\ww}}}, & ... & a_{M,K}\bits{{\ww}{-}{1}{:}{\Hw{\ww}}} \end{bmatrix}$
    \State $\bld{A_0} = \begin{bmatrix} a_{1,1}\bits{{\Hw{\ww}}{-}{1}{:}{0}}, & ... & a_{1,K}\bits{{\Hw{\ww}}{-}{1}{:}{0}} \\ ... & ... & ... \\ a_{M,1}\bits{{\Hw{\ww}}{-}{1}{:}{0}}, & ... & a_{M,K}\bits{{\Hw{\ww}}{-}{1}{:}{0}} \end{bmatrix}$
    \State $\bld{B_1} = \begin{bmatrix} b_{1,1}\bits{{\ww}{-}{1}{:}{\Hw{\ww}}}, & ... & b_{1,N}\bits{{\ww}{-}{1}{:}{\Hw{\ww}}} \\ ... & ... & ... \\ b_{K,1}\bits{{\ww}{-}{1}{:}{\Hw{\ww}}}, & ... & b_{K,N}\bits{{\ww}{-}{1}{:}{\Hw{\ww}}} \end{bmatrix}$
    \State $\bld{B_0} = \begin{bmatrix} b_{1,1}\bits{{\Hw{\ww}}{-}{1}{:}{0}}, & ... & b_{1,N}\bits{{\Hw{\ww}}{-}{1}{:}{0}} \\ ... & ... & ... \\ b_{K,1}\bits{{\Hw{\ww}}{-}{1}{:}{0}}, & ... & b_{K,N}\bits{{\Hw{\ww}}{-}{1}{:}{0}} \end{bmatrix}$
    \vspace*{0.15cm}
    \State $\bld{A_s} = \bld{A_1}+\bld{A_0}$ \label{kmm:line:kmm-add-a}
    \State $\bld{B_s} = \bld{B_1}+\bld{B_0}$ \label{kmm:line:kmm-add-b}
    \State $\bld{C_1} = \alg.\KMM'*?{n/2}/{\Hwf{\ww}}\lb \bld{A_1},\bld{B_1} \rb$
    \State $\bld{C_s} = \alg.\KMM'*?{n/2}/{{\Hw{\ww}}{+}{1}}\lb \bld{A_s},\bld{B_s} \rb$
    \State $\bld{C_0} = \alg.\KMM'*?{n/2}/{\Hw{\ww}}\lb \bld{A_0},\bld{B_0} \rb$
    \State $\bld{C} = \bld{C_1} \ll \ww$ \label{kmm:line:kmm-output-start}
    \State $\bld{C} \mathrel{+}= \lb \bld{C_s} - \bld{C_1} - \bld{C_0} \rb\ll\Hw{\ww}$ \label{kmm:line:kmm-add-c1}
    \State $\bld{C} \mathrel{+}= \bld{C_0}$ \label{kmm:line:kmm-add-c2}
     \label{kmm:line:kmm-output-end}
    \Else
    \State $\bld{C} = \alg.\MM'*?\one/\ww \lb \bld{A}, \bld{B} \rb$ \label{kmm:line:kmm-mm1}
    \EndIf
    \State return $\bld{C}$
    \EndFunction
  \end{algorithmic}
\end{algorithm}

\fig{kmm:fig:kmm} shows the 2-digit Karatsuba scalar multiplication algorithm
\cite{karatsuba1962} from \fig{kmm:fig:ksm} extended to matrix multiplication
analogously to how \fig{kmm:fig:mmn} extends conventional 2-digit scalar
multiplication in \fig{kmm:fig:sm} to matrix multiplication.
Algorithm \ref{kmm:alg:kmm} shows the generalization of this, where n-digit
Karatsuba matrix multiplication is performed by carrying out the same steps
recursively for each smaller-bit matrix multiplication.
In Algorithm \ref{kmm:alg:kmm}, the full matrix product is split into three
separate partial-product matrix multiplications between matrices each containing
bit slices of every element.
The elements in matrices $\bld{A_0}$ and $\bld{B_0}$ contain the lower bits
(bits ${\hW>\ww}{-}{1}$ down to 0) of every element in the $\bld{A}$ and
$\bld{B}$ matrices, while $\bld{A_1}$ and $\bld{B_1}$ contain the upper bits
(bits ${\ww}{-}{1}$ down to $\hW>\ww$) of every element in matrices $\bld{A}$
and $\bld{B}$. The $\bld{A_s}$ and $\bld{B_s}$ matrices are formed by summing
$\bld{A_1} + \bld{A_0}$ and $\bld{B_1} + \bld{B_0}$, and therefore their
elements have a bitwidth of $\Hw{\ww}+1$.
The partial-product matrices are then summed analogously to how the partial
scalar products are summed after multiplication in \ksm from Algorithm
\ref{kmm:alg:ksm}.

\subsection{\kmm Complexity Analysis}
\label{kmm:sec:alg-c}
In this subsection, we derive the complexity of \kmm and compare it to the
complexity of the conventional \mmn, and \ksm algorithms. To do this, we
decompose each algorithms' complexity to number of $\ww$-bit multiplications,
additions, and shift operations.
This provides a general technology-agnostic foundation for evaluating different
possible \kmm hardware implementations and modelling the costs and benefits of
implementing the algorithm in hardware across different possible implementation
technologies where the cost of each type of operation may vary depending on the
implementation platform used. For example, implementations on FPGA may result in
multipliers mapping to DSP units, additions and accumulations mapping to soft
look-up-table (LUT) and register resources, whereas ASIC implementations will
result in different costs and trade-offs than this for each type of operation.

\begin{figure}[]
  \vspace*{-0.1cm}
  \centering
  \includegraphics[scale=0.0525]{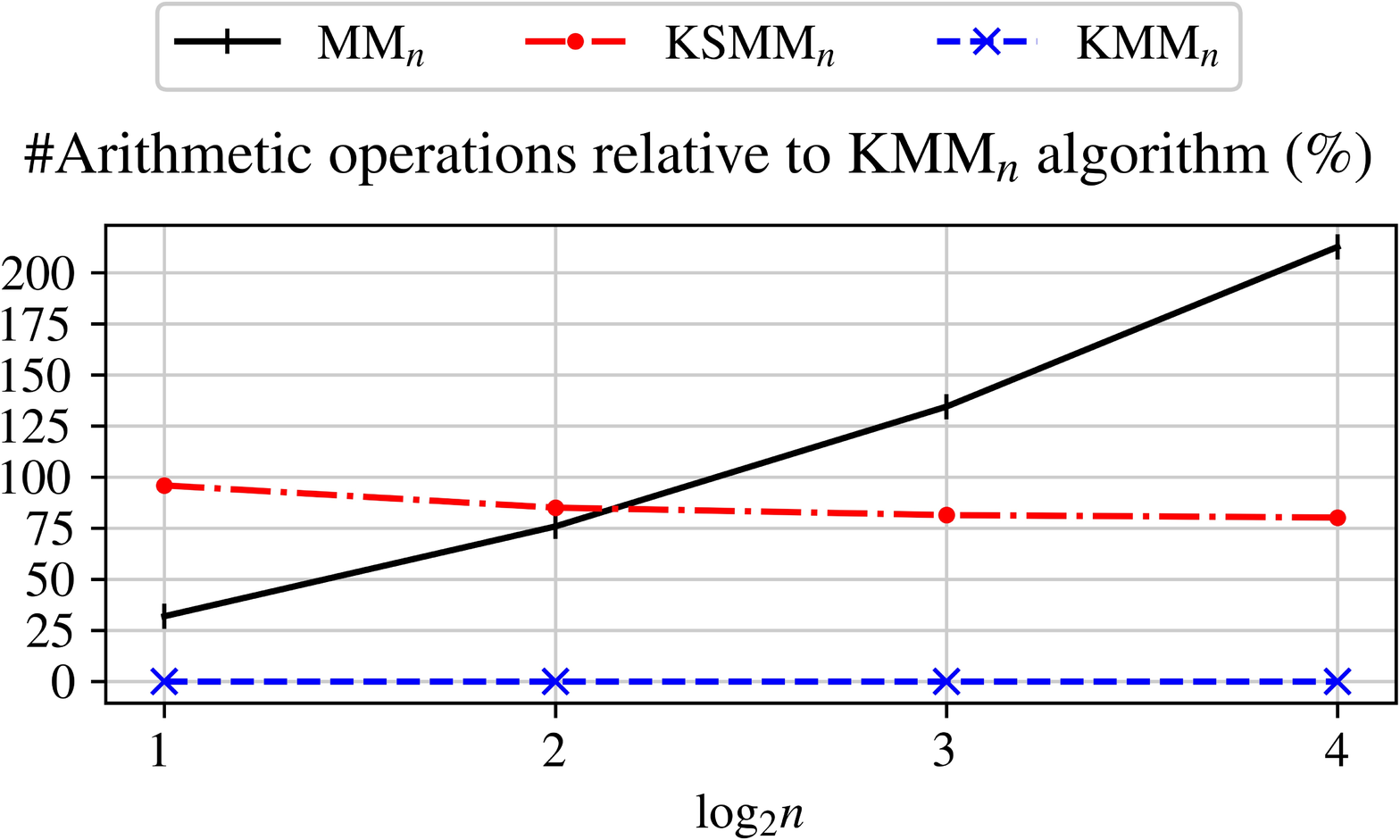}
  \vspace*{-0.6cm}
  \caption{Plotting \eq{kmm:eq:mm-c} and \eq{kmm:eq:ksm-c} relative to
  \eq{kmm:eq:km-c} for different $\nn$ with $\dd = 64$. As can be seen, \ksmm?\nn
  requires over 75\% more operations than \kmm?\nn.
  Additionally, \kmm?\nn and \ksmm?\nn require
  exponentially fewer operations than \mmn?\nn with respect to $\nn$, however,
  \kmm?\nn requires fewer operations than \mmn?\nn even starting at $n = 2$,
  while \ksmm?\nn does not fall below \mmn?\nn until $n > 4$.
  }
  \label{kmm:fig:arith-c}
\end{figure}

Additionally, while the main focus of this work is on leveraging \kmm in custom
hardware designs, we also compare \kmm's complexity more simply in number of
arithmetic operations to allow modelling the time complexity of
\kmm execution on general-purpose hardware containing fixed operator word sizes.
This analysis (plotted in \fig{kmm:fig:arith-c}) indicates that
\kmm requires significantly fewer operations to
execute large-integer matrix multiplication on general-purpose hardware than
conventional \ksm or \mmn algorithms. This is relevant when
the matrix element bitwidths are larger than the word size
of the general-purpose hardware operators, for example, inputs larger
than 32 bits when executing on a CPU containing arithmetic logic units (ALU)s
that support 32-bit inputs.

\subsubsection{\mmn Complexity}
The complexity of conventional \ndigit \mmn between two matrices of size
$d\times d$ is derived by counting the number of
operations that are performed in Algorithm \ref{kmm:alg:mmn}:
\begin{subequations}
 \label{kmm:eq:mmn-all-alg-c}
\begin{align} \label{kmm:eq:mmn-alg-c}
    \cc{\alg.\MMN'*?\nn/\ww} =&\spc \cc{\alg.\MMN'*?{n/2}/{\Hwf{\ww}}}
    + 3\spc \cc{\alg.\MMN'*?{n/2}/{\Hw{\ww}}}
    \notag\\ &+\dd^2 \lba\opcc{\add{\ww{+}\wa}} + 2\spc\opcc{\add{2w{+}\wa}}\rba
    \notag\\ &+\dd^2 \lba\opcc{\shift{\ww}} + \opcc{\shift{\Hw{\ww}}}\rba\\
    \label{kmm:mm1-c}
    \cc{\alg.\MM'*?\one/\ww} =&\spc \dd^3\lba\opcc{\mult{\ww}} + \opcc{\accum{2\ww}}\rba
    \,.
\end{align} \end{subequations}
Typically, $\opcc{\accum{2\ww}} = \opcc{\add{2\ww+\wa}}$, where $\wa$ is an
additional bitwidth added to account for accumulation.
However, in \secn{kmm:sec:accum-reduction}, we discuss a method for reducing
the complexity of the accumulations to be less than this.

The $\add{\ww{+}\wa}$ terms in \eq{kmm:eq:mmn-alg-c} come from the additions
forming the $\lb\bld{C_{10}} + \bld{C_{01}}\rb$ term on line
\ref{kmm:line:mmn-add1} of
Algorithm \ref{kmm:alg:mmn}. Here, the bitwidth of the $\bld{C_{10}}$ and
$\bld{C_{01}}$ elements is $\ww+\wa$ because they are accumulations of $\ww$-bit
products of ${\Hwf{\ww}}$ and ${\Hw{\ww}}$-bit values.
The two $\add{2w{+}\wa}$ terms in \eq{kmm:eq:mmn-alg-c} come from the additions to
$\bld{C}$ on lines \ref{kmm:line:mmn-add1} and \ref{kmm:line:mmn-add2} of Algorithm
\ref{kmm:alg:mmn}. The bitwidth of these additions is kept on $2w+\wa$ bits since
$\bld{C}$ results in accumulations of ${2}{\ww}$-bit products of $\ww$-bit
values.

\subsubsection{\ksm Complexity}
The complexity of \ksm is derived by counting the operations performed in
Algorithm \ref{kmm:alg:ksm}:
\begin{subequations} \label{kmm:eq:ksm-alg-c} \begin{align}
 \label{kmm:eq:ksmn-alg-c}
    \cc{\alg.\KSM'*?\nn/\ww} =
    2\spc&\lba \opcc{\add{2w} +
    \opcc{\add{\Hw{\ww}}} + \opcc{\add{2\Hw{\ww}{+}{4}}}\rba
    \notag\\  +\spc& \opcc{\shift{\ww}} + \opcc{\shift{\Hw{\ww}}}}
    \notag\\ +\spc&\cc{\alg.\KSM'*?{n/2}/{\Hwf{\ww}}}
    +\cc{\alg.\KSM'*?{n/2}/{\Hw{\ww}{+}{1}}}
    \notag\\+\spc&\cc{\alg.\KSM'*?{n/2}/{\Hw{\ww}}}\\
    \label{kmm:eq:ksm1-alg-c}
    \cc{\alg.\KSM'*?1/\ww} = \spcc \opcc{&\mult{\ww}}
    \,.
\end{align} \end{subequations}

The two $\add{{\Hw{\ww}}}$ terms in \eq{kmm:eq:ksmn-alg-c} come from the
$\Hw{\ww}$-bit additions forming the $a_{s}$ and $b_{s}$ terms on lines
\ref{kmm:line:ksm-add-a} and \ref{kmm:line:ksm-add-b} of Algorithm
\ref{kmm:alg:ksm}.
The two $\add{2\Hw{\ww}{+}4}$ terms in \eq{kmm:eq:ksmn-alg-c} come from forming the
$\lb c_{s} - c_{1} - c_{0} \rb$ term on line \ref{kmm:line:ksm-add-c1} of Algorithm
\ref{kmm:alg:ksm}, where these terms can be first summed together on $2\Hw{\ww}+4$
bits before being shifted and added to the other product terms. The bitwidth
$2\Hw{\ww}+4$ is required because $c_{s}$ is a $(2\Hw{\ww}{+}2)$-bit product of
$(\Hw{\ww}{+}1)$-bit values, and the additional two bits are to account for sign
extension and subtraction of the $c_1$ and $c_0$ terms.
The two $\add{2w}$ terms in \eq{kmm:eq:ksmn-alg-c} come from the additions to
$c$ on lines \ref{kmm:line:ksm-add-c1} and \ref{kmm:line:ksm-add-c2} of
Algorithm \ref{kmm:alg:ksm}. These additions are on $2w$-bit values since $c$
will ultimately result in the ${2}{\ww}$-bit product of two $\ww$-bit values.

\subsubsection{\ksmm Complexity}
To compare \ksm to \kmm and the other matrix multiplication algorithms, we
analyze the complexity of an algorithm we refer to as \ksmm. \ksmm is defined as
a conventional matrix multiplication algorithm as in \eq{eq:MM}, but where \ksm is used for the multiplications between all elements rather than conventional scalar
multiplication.
\ksmm then has the following complexity:
\begin{align}
\label{kmm:eq:ksMM-alg-c}
    \cc{\alg.\ksMM'*?\nn/\ww} = \dd^3\lba \cc{\alg.\KSM'*?\nn/\ww}
    + \opcc{\accum{2\ww}} \rba
    \,.
\end{align}

\subsubsection{\kmm Complexity}
The complexity of \kmm is derived by counting the operations performed in
Algorithm \ref{kmm:alg:kmm}:
\begin{subequations} \label{kmm:eq:kmm-alg-c} \begin{align}
 \label{kmm:eq:kmmn-alg-c}
    \cc{\alg.\KMM'*?\nn/\ww} =
    \spc&2\spc \dd^2\lba\opcc{\add{2\Hw{\ww}{+}{4}{+}{\wa}}} + \spc \opcc{\add{2\ww{+}{\wa}}}\rba
    \notag\\ +\spc& \dd^2\lba 2\spc\opcc{\add{\Hw{\ww}}} +\opcc{\shift{\ww}} + \opcc{\shift{\Hw{\ww}}}\rba
    \notag\\ + \spc&\cc{\alg.\KMM'*?{n/2}/{\Hwf{\ww}}} + \cc{\alg.\KMM'*?{n/2}/{\Hw{\ww}{+}{1}}}
    \notag\\ +\spc &\cc{\alg.\KMM'*?{n/2}/{\Hw{\ww}}}\\ \label{kmm:eq:kmm1-alg-c}
     \cc{\alg.\KMM'*?1/\ww} &= \cc{\alg.\MM'*?\one/\ww}
    \,.
\end{align} \end{subequations}

The two $\add{{\Hw{\ww}}}$ terms in \eq{kmm:eq:kmmn-alg-c} come from the
$\Hw{\ww}$-bit additions forming the $\bld{A_s}$ and $\bld{B_s}$ terms on lines
\ref{kmm:line:kmm-add-a} and \ref{kmm:line:kmm-add-b} of Algorithm
\ref{kmm:alg:kmm}.
The two $\add{2\Hw{\ww}{+}{4}{+}\wa}$ terms in \eq{kmm:eq:kmmn-alg-c} come from
forming the $\lb \bld{C_s} - \bld{C_1} - \bld{C_0} \rb$ term on line
\ref{kmm:line:kmm-add-c1} of Algorithm \ref{kmm:alg:kmm}, where these terms can
be first
summed together on $2\Hw{\ww}+4+\wa$ bits before being shifted and added to the
other product terms. The bitwidth $2\Hw{\ww}+4+\wa$ is required because the
bitwidth of $\bld{C_s}$ is $2\Hw{\ww}+2+\wa$ since it is accumulations of
$(2\Hw{\ww}{+}2)$-bit products of $(\Hw{\ww}{+}1)$-bit values, and the
additional two bits are to account for sign extension and subtraction of the
$\bld{C_1}$ and $\bld{C_0}$ terms.
The two $\add{2w{+}\wa}$ terms in \eq{kmm:eq:kmmn-alg-c} come from the additions to
$\bld{C}$ on lines \ref{kmm:line:kmm-add-c1} and \ref{kmm:line:kmm-add-c2} of
Algorithm \ref{kmm:alg:kmm}. The bitwidth of these additions is kept on $2w+\wa$
bits since $\bld{C}$ results in accumulations of ${2}{\ww}$-bit products of
$\ww$-bit values.

\eq{kmm:eq:kmmn-alg-c} shows that \kmm significantly reduces the complexity of
the 8 addition and shift operations in \eq{kmm:eq:ksmn-alg-c} that are performed
$(n/2)\klog\dd^3$ times in \ksmm by reducing their occurrence by a factor of
$\dd$.
On the other hand, \kmm trades $\dd^3$ accumulations of $2\ww$-bit values
in \eqs (\ref{kmm:mm1-c}) or (\ref{kmm:eq:ksMM-alg-c}) for
$n\klog\dd^3$ smaller-width accumulations in \eq{kmm:eq:kmm1-alg-c}.
However, in \secn{kmm:sec:accum-reduction} we show how the penalty of this in
hardware is mitigated when combining \kmm with an alternative accumulation
algorithm.

\subsubsection{Arithmetic complexity}
If only counting the number of operations without considering operation
bitwidths or type, we can simplify
\eq{kmm:eq:mmn-all-alg-c} to:
\begin{align}
\label{kmm:eq:mm-c}
    \cc{\alg.\MMN'*?\nn/\ww<{}} =&\spc
    2\spc n^2 \dd^3 + 5\spc (n/2)^2 \dd^2
    \,,
\end{align}
\eq{kmm:eq:ksMM-alg-c} can be simplified to:
\begin{align}
\label{kmm:eq:ksm-c}
    \cc{\alg.\ksMM'*?\nn/\ww<{}} =&\spc
    \lb 1 + 11\spc(n/2)\klog\rb \spc\dd^3
    \,,
\end{align}
and \eq{kmm:eq:kmm-alg-c} can be simplified to:
\begin{align}
\label{kmm:eq:km-c}
    \cc{\alg.\KMM'*?\nn/\ww<{}} =&\spc
    (n/2)\klog \lb 6\spc \dd^3 + 8\spc\dd^2 \rb
    \,.
\end{align}

\subsection{Mitigating the Accumulator Complexity Increase in \kmm}
\label{kmm:sec:accum-reduction}
As found in \secn{kmm:sec:alg-c}, \kmm has one penalty of trading $\dd^3$
accumulations of $2\ww$-bit values in \eqs (\ref{kmm:mm1-c}) or
(\ref{kmm:eq:ksMM-alg-c}) for $n\klog\dd^3$ smaller-width accumulations in
\eq{kmm:eq:kmm1-alg-c}. In this subsection, we show how this downside is mitigated
when using Algorithm \ref{kmm:alg:mm1} as the \mm algorithm in \kmm on line
\ref{kmm:line:kmm-mm1} of Algorithm \ref{kmm:alg:kmm}. Algorithm
\ref{kmm:alg:mm1} performs \mm using an alternative accumulation structure that
reduces the accumulation hardware complexity.

\begin{algorithm}[]
  \algsize
  \caption{MM$_\one$ algorithm with reduced accumulator complexity used in the
  baseline \fpMm MXUs of all compared architectures. $p$ is defined as the
  number of multiplication products that are pre-accumulated on a smaller
  bitwidth to reduce the accumulation complexity before being added to the
  full-bitwidth accumulation sum. We use $p = 4$ in our evaluation.}
  \label{kmm:alg:mm1}
  \begin{algorithmic}[1]
    \Function{\alg.\MM'?\one<{}}{$\bld{A}$, $\bld{B}$, $p$}
    \For {$i$ = 0; $i <M$; $i$ ++}
    \For {$j$ = 0; $j <N$; $j$ ++}
    \State $\bld{C}_{i,j} = 0$
    \For {$k$ = 0; $k <K$; $k \mathrel{+}= p$}
    \State $x = 0$
    \For {$q$ = 0; $q < p$; $q$ ++}
    \State $x \mathrel{+}= \bld{A}_{i,k+q} \times \bld{B}_{k+q,j} $
    \EndFor
    \State $\bld{C}_{i,j} \mathrel{+}= x$
    \EndFor
    \EndFor
    \EndFor
    \State return $\bld{C}$
    \EndFunction
  \end{algorithmic}
\end{algorithm}

In conventional matrix multiplication, each product of $\ww$-bit elements is
added to a running sum kept on $2\ww+\wa$ bits,
where $\wa = \ceil{\tx{log}_2\dd}$ and is an extra bitwidth
added to account for accumulation in order to accumulate $\dd$ elements which
adds extra hardware complexity.
This means that normally $p$ accumulations of $2\ww$-bit elements will require
being added to a $(2\ww+\wa)$-bit running sum and each addition will be on
$2\ww+\wa$ bits and therefore contain the following complexity:
\begin{align}
    p\spc\opcc{\accum{2\ww}} = p\spc\opcc{\add{2\ww+\wa}}
    \,.
\end{align}
However, the average bitwidth of the addition operations is reduced when using
Algorithm \ref{kmm:alg:mm1} for accumulation of $p$ elements of
bitwidth $2\ww$ because $p$ elements are first added together in isolation on a
smaller running sum requiring a bitwidth of only $2\ww+\wP$ bits for keeping $p$
elements, where $\wP = \ceil{\tx{log}_2p}$. Only after this initial pre-sum will
this result then be
added to the full running sum that is kept on
a larger $2\ww+\wa$ bits for keeping $\dd$ elements. This reduces the
average bitwidth for every $p$ accumulations to the following:
\begin{align}
    p\spc\opcc{\accum{2\ww}} =&\spc \opcc{\add{2\ww+\wa}}
    + (p-1)\spc\opcc{\add{2\ww+\wP}}
    \,.
\end{align}
Furthermore, in systolic-array architectures, each accumulation output is
buffered in a dedicated register, which adds further hardware complexity to the
accumulation operation. However, the number of required accumulation registers
when using Algorithm \ref{kmm:alg:mm1} is also reduced by a factor of $p$ as shown
in the hardware implementation from \fig{kmm:pes} in \sections \ref{kmm:arch-base}
since the accumulation result only needs to be buffered after being added to the
full running sum kept on $2\ww+\wa$ bits.

\section{\kmm Hardware Architectures}
\label{kmm:sec:arch}

In this section, we present a general family of hardware architectures for
efficiently exploiting the \kmm algorithm in hardware and derive metrics for
analyzing the area or execution time benefits of the \kmm architectures.
The first type of \kmm architecture, described in \secn{kmm:sec:fp-kmm}, is
a fixed-precision architecture optimized for executing inputs that are not
expected to vary in bitwidth. We then present a precision-scalable \kmm
architecture in \secn{kmm:sec:ps-kmm} that can more efficiently execute across
multiple input precisions for applications where the input bitwidths are
expected to vary.

\subsection{Baseline \fpMm Architecture}
\label{kmm:arch-base}

\begin{figure}[]
  \centering
  \includegraphics[scale=.9]{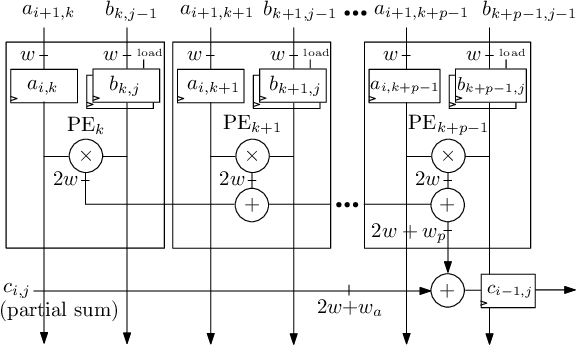}
  \vspace*{-0.5cm}
  \caption{Showing the internal PE structure of the \fpMm MXUs shown in
    \fig{kmm:fig:MM-mxu} as well as the structure for
    implementing Algorithm \ref{kmm:alg:mm1} in hardware to reduce the hardware cost
    of the accumulator logic. $p$ is a hardware parameter equal to the number of
    multiplication products that are pre-accumulated on a smaller bitwidth to
    reduce the accumulation complexity before being added to the full-bitwidth
    accumulation sum. We use $p = 4$ in our evaluation.}
  \label{kmm:pes}
\end{figure}

\begin{figure}[]
  \centering
  \includegraphics[scale=1.35]{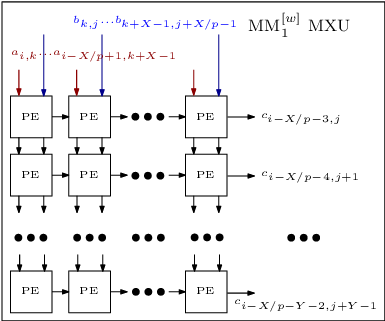}
  \vspace*{-0.5cm}
  \caption{Baseline \fpMm MXU architecture present at the core of the \kmmArch
    architectures, provided for context.
  $X$ and $Y$ refer to the MXU width and height in number of multipliers.
  }
  \label{kmm:fig:MM-mxu}
\end{figure}

\Figure \ref{kmm:fig:MM-mxu} shows the internal structure of each baseline \fpMm MXU
at the core of each \kmmArch architecture, and \fig{kmm:pes} shows the internal
structure of the processing elements (PE)s inside the \fpMm MXUs. \fig{kmm:pes} also
shows the structure for how Algorithm \ref{kmm:alg:mm1} from
\secn{kmm:sec:accum-reduction}
can be implemented in hardware and how the algorithm
is able to reduce the hardware cost of the accumulator logic. This accumulation
structure allows for the number of $({2w}{+}{\wa})$-bit accumulation adders and
their output registers to be reduced by a factor of $p$, where they are instead
traded for additions on lower-bitwidth values in the range of $2\ww$ to
$2\ww+\ceil{\tx{log}_2p}$ bits that do not require their output to be buffered in
registers.

\subsection{Fixed-Precision \fpKmm Architecture}
\label{kmm:sec:fp-kmm}

\begin{figure}[]
  \centering
  \includegraphics[scale=.9]{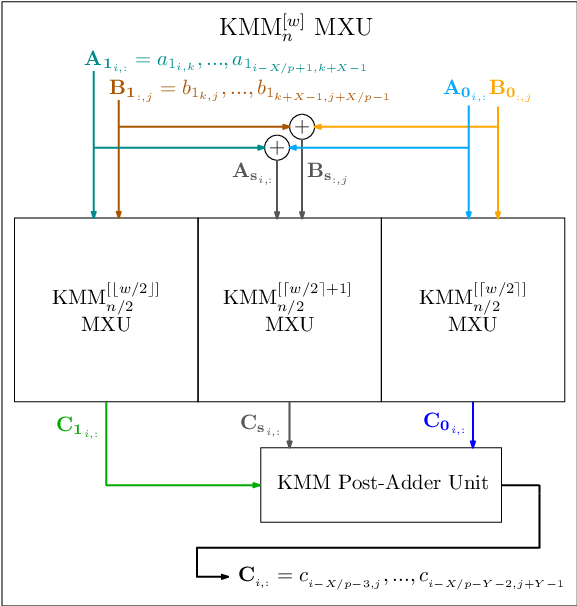}
  \vspace*{-0.5cm}
  \caption{Fixed-precision \fpKmm architecture for executing on inputs of a
    fixed precision of $\win$ bits.}
  \label{kmm:fig:kmm-mxu}
\end{figure}

\Figure \ref{kmm:fig:kmm-mxu} shows the proposed fixed-precision \fpKmm
architecture for executing on inputs of a fixed precision of $\win$ bits that
are not expected to vary in bitwidth.
Rather than having one MXU with $\win$-bit-input multiplier units, this
architecture consists of three sub-MXUs that compute matrix multiplication on
either $\Hwf{\win}$, $\Hw{\win}{+}1$, or $\Hw{\win}$-bit inputs.

The additions on lines \ref{kmm:line:kmm-add-a} and \ref{kmm:line:kmm-add-b} of
Algorithm \ref{kmm:alg:kmm} are performed on $X$ scalar adders at the MXU inputs.
Similarly, the additions on lines \ref{kmm:line:kmm-add-c1} and
\ref{kmm:line:kmm-add-c2} of Algorithm \ref{kmm:alg:kmm} are performed on $Y$
scalar adders at the MXU outputs. Due to the nature of right/left shifting by a
constant offset in custom hardware, the shift operations at the output of the
MXUs do not require any area overhead.
If desired, each of the three sub-MXUs can also be instantiated as another
\fpKmm MXU containing three more sub-MXUs to implement additional levels of \kmm
recursion. The final level of MXUs will be \fpMm MXUs.

\begin{figure}[]
  \centering
  \includegraphics[scale=.9]{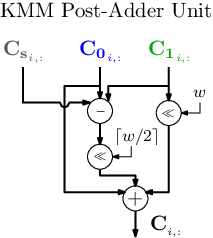}
  \vspace*{-0.1cm}
  \caption{KMM Post-Adder Unit from \fig{kmm:fig:kmm-mxu} for executing
    $\mathbf{C_1}_{_{i,:}}{\ll}{w} + \left(\mathbf{C_s}_{_{i,:}} - \mathbf{C_1}_{_{i,:}} - \mathbf{C_0}_{_{i,:}} \right){\ll}{\lceil} {w}{/}{2}{\rceil} + \mathbf{C_0}_{_{i,:}}$.}
  \label{kmm:fig:kmm-add-u}
\end{figure}

\subsection{Precision-Scalable \psKmm Architecture}
\label{kmm:sec:ps-arch}
\label{kmm:sec:ps-kmm}

\begin{figure}[]
  \centering
  \includegraphics[scale=0.9]{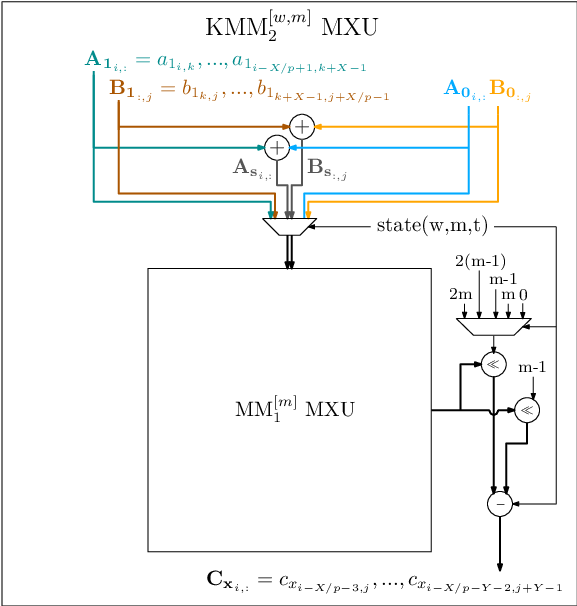}
  \vspace*{-0.5cm}
  \caption{Precision-scalable \psKmm architecture for more efficiently using
    $\wm$-bit-input multipliers to execute across varying input precisions of
    bitwidth $\win$ for applications where the input bitwidths are expected to
    vary.}
  \label{kmm:fig:kmm-ps-single-mxu}
\end{figure}

\Figure \ref{kmm:fig:kmm-ps-single-mxu} shows the proposed precision-scalable \psKmm
architecture for implementing one level of \kmmArch recursion. This architecture
can more efficiently use $\wm$-bit-input multipliers to execute across varying
input precisions of bitwidth $\win$ for applications where the input bitwidths
are expected to vary.
Unlike in prior works \citePsML, the minimum possible execution time when fully
utilizing the compute resources scales less than quadratically with the input
bitwidths.
As discussed further in \secn{kmm:sec:system}, the input matrices are divided
into tiles and fed into the MXU one-by-one to perform GEMM.
In this architecture, each set of input matrix tiles may be read multiple times
and either the \mm, \mmn?2, or \kmm?2 algorithm may be executed depending on the
input bitwidths $\win$ and the multiplier bitwidth $\wm$. An iteration state
signal $t$ is reset when a new set of input tiles is read and is incremented
each time the same set of input tiles is re-read.

\subsubsection{\mm and \mmn?2 Mode}
\label{kmm:sec:kmm-arch-mm-mode}
If $\win \le \wm$, the architecture will execute the \fpMm/\ww algorithm,
bypassing any MXU input/output addition or shifting steps, $\bld{A_0}$ and
$\bld{B_0}$ will be fed into the MXU as inputs, and each set of input tiles is
read only once.

If $2m-2 < \win \le 2\wm$, the architecture will execute the \psMm?2/\ww algorithm
and each set of input matrix tiles will be read a total of four times before
proceeding to the next set of input tiles. The \mmn?2 algorithm is used instead
of \kmm?2 for this input bitwidth range because the bitwidth of the elements in
the $\bld{A_s}$ and $\bld{B_s}$ matrices in Algorithm \ref{kmm:alg:kmm} would be
$\wm+1$ which would be too large by 1 bit to fit onto the $\wm$-bit multipliers
in the MXU.
In each read for this input bitwidth range, the MXU will accept either the
$\bld{A_1}$ and $\bld{B_1}$ inputs or the $\bld{A_0}$ and $\bld{B_0}$ inputs
depending on the tile read iteration $t$.
$\bld{A_1}$ and $\bld{B_1}$ will contain bits $2\wm-1$ down to $\wm$ of the
$\bld{A}$ and $\bld{B}$ matrix elements.
$\bld{A_0}$ and $\bld{B_0}$ will contain bits $\wm-1$ down to $0$ of the
$\bld{A}$ and $\bld{B}$ matrix elements.

The MXU output vectors $\bld{C_x}_{_{i,:}}$ in \fig{kmm:fig:kmm-ps-single-mxu} will
be equal to either $(\bld{C_1 }_{_{i,:}}\ll 2\wm)$, $(\bld{C_{10}}_{_{i,:}} \ll \wm)$,
$(\bld{C_{01}}_{_{i,:}} \ll \wm)$, or $\bld{C_0 }_{_{i,:}}$ depending on the tile read iteration $t$
to incrementally execute lines
\ref{kmm:line:mmn-output-start}-\ref{kmm:line:mmn-output-end} of Algorithm \ref{kmm:alg:mmn}
throughout the tile read iterations,
where $m$ is considered equivalent to the value of $\Hw{\ww}$ in Algorithm \ref{kmm:alg:mmn}.
Specifically, depending on the tile read iteration $t$, the MXU output vectors
will be equal to $(\bld{C_1 }_{_{i,:}}\ll 2\wm)$ to form the addition on line
\ref{kmm:line:mmn-output-start} of Algorithm \ref{kmm:alg:mmn}, $\bld{C_0 }_{_{i,:}}$ to
form the addition on line \ref{kmm:line:mmn-output-end},
and separately $(\bld{C_{10}}_{_{i,:}} \ll \wm)$ or $(\bld{C_{01}}_{_{i,:}} \ll \wm)$
to collectively form the addition on line \ref{kmm:line:mmn-add1}.

Each partial matrix tile product will need to be accumulated with prior ones
outside of the MXU, however, this is the same functionality already present in
GEMM where multiple matrix tile products must be summed to form a final matrix
product, and this functionality will therefore already be present in GEMM
accelerators outside of the MXU such as in the GEMM and ML accelerator system
from our prior work \cite{pogue2024fast}.

\subsubsection{\kmm?2 Mode}
If $\wm < \win \le 2m-2$, the architecture will execute the \fpKmm?2/\ww algorithm
and each set of input matrix tiles will be read a total of three times before
proceeding to the next set of input tiles. For each read, the MXU will accept or
form either the $\bld{A_1}$ and $\bld{B_1}$ inputs, the $\bld{A_s}$ and
$\bld{B_s}$ inputs, or the $\bld{A_0}$ and $\bld{B_0}$ inputs depending on the
tile read iteration $t$.
$\bld{A_1}$ and $\bld{B_1}$ will contain bits $2(\wm-1)-1$ down to $\wm-1$ of
the $\bld{A}$ and $\bld{B}$ matrix elements.
$\bld{A_0}$ and $\bld{B_0}$ will contain bits $\wm-2$ down to $0$ of the
$\bld{A}$ and $\bld{B}$ matrix elements.
The MXU output vectors $\bld{C_x}_{_{i,:}}$ in \fig{kmm:fig:kmm-ps-single-mxu} will
be equal to either
$\left[\lb  \bld{C_1}_{_{i,:}}\ll 2(\wm-1)\rb -
  \lb\bld{C_1}_{_{i,:}} \ll (\wm-1)\rb\right]$, $\left[\bld{C_s}_{_{i,:}} \ll
  (\wm-1)\right]$, or $\left[\bld{C_0}_{_{i,:}} - \lb\bld{C_0}_{_{i,:}} \ll (\wm-1)\rb\right]$ depending on the tile read iteration $t$
to incrementally execute lines
\ref{kmm:line:kmm-output-start}-\ref{kmm:line:kmm-output-end} of Algorithm \ref{kmm:alg:kmm}
throughout the tile read iterations, where $m-1$ is considered equivalent to the
value of $\Hw{\ww}$ in Algorithm \ref{kmm:alg:kmm}.

Each partial matrix tile product will need to be accumulated with prior ones
outside of the MXU, however, this functionality will already be present in GEMM
accelerators as explained above in \secn{kmm:sec:kmm-arch-mm-mode}.

A precision-scalable \psMm?2 architecture can also be implemented that has a
similar structure as the precision-scalable \psKmm architecture, except that it
will only either execute the \fpMm/\ww algorithm if $\win \le \wm$ or the
\psMm?2/\ww algorithm if $\wm < \win \le 2m$.
We also note that a precision-scalable \psKsmm architecture exploiting \ksm?2
would not be as efficient to implement in hardware compared to a
precision-scalable \psKmm architecture. This is because, in addition to the
extra adders that would be required at the output/inputs of every multiplier as
discussed in \secn{kmm:sec:alg-c}, multiplexers would also have to be placed at
the output/inputs of every multiplier in the MXU as well for output/input
arbitration depending on the width of the inputs. In contrast, the \psKmm
architecture reduces this extra adder complexity as already discussed, and it
can employ an efficient more conventional systolic array at the core not
requiring multiplexers surrounding each multiplier.

\subsection{System Integration}
\label{kmm:sec:system}
In order to perform GEMM on an MXU and multiply matrices of arbitrary sizes that
can be larger than the MXU dimensions, the input matrices are divided into tiles
and fed to the MXU one-by-one. Following each tile multiplication, the partial
tile products are accumulated outside of the MXU to generate each final matrix
product tile. Prior to each tile multiplication, a $\bld{B}$ tile is loaded into
the MXU. It then remains in place as the $\bld{A}$ tile flows through the MXU
producing the tile product, during which a new $\bld{A}_{i,:}$ vector is fed
into the MXU each clock cycle. Additionally, to hide the latency of loading
$\bld{B}$ tiles, the MXU PEs each contain one extra $b$ buffer to load the next
$\bld{B}$ tile into the MXU as the current tile is being multiplied, where each
extra $b$ buffer in the PEs will hold one individual element of the next
$\bld{B}$ tile after it is loaded.

The presented \kmmArch architectures are illustrated for unsigned integer
inputs, however, if the inputs are signed, a 1-dimensional adder vector can be
used to add a constant offset to the inputs of an MXU to convert them to
unsigned. The zero-point adjuster method from our previous work
\cite{pogue2024fast} can then be used to efficiently eliminate the effects of
this constant offset in the matrix products before exiting the MXU.

We use an ML accelerator system design based on the one from our previous work
\cite{pogue2024fast}, which has open-source code available \cite{ffip-source},
to house and evaluate the \kmmArch and baseline MXU architectures.
We were able to swap the precision-scalable \psKmm MXU architecture from
\fg{kmm:fig:kmm-mxu} into our system design \cite{pogue2024fast} in place of the
free-pipeline fast inner-product (FFIP) MXU.
This change was mostly seamless but also required updates to the memory system such
that each set of input matrix tiles can optionally be re-read up to three or
four times before proceeding to the next set of input tiles. The number of times
that the matrix tiles are re-read and the purpose for this is explained in
\secn{kmm:sec:ps-arch}.

\subsection{Multiplier Compute Efficiency}
\label{kmm:sec:mu}
In this subsection, we define a performance-per-area metric called the
multiplier compute efficiency in \eq{kmm:mu} which we use to compare the \psKmm
architecture against baseline designs and prior works. The metric is used to
compare the amount of computational work that can be performed per compute area
regardless of the clock frequency or input bitwidths.
The importance of this property is expanded upon more later in this subsection,
as well as in \secn{kmm:metrics}.

The hardware complexity of fixed-point multipliers typically scale
\quadratically with the input bitwidth compared to linearly for adders and
registers \multComplexity, causing the hardware footprint of multipliers to
dominate that of adders and registers. Due to this, multipliers and MAC units
are commonly the area-dominant computational resources in deep learning and
GEMM-based accelerators \multDominant.
Therefore, we derive a performance-per-area metric defined below for quantifying
how much the algebraic optimizations exploited in an architecture reduce the
computational complexity of the area-dominant operations (multiplications) and
measure how effectively an architecture can utilize these
resources relative to a conventional design using no
algebraic optimizations:
\begin{align}  \label{kmm:in-mu}
  \frac{\tx{mults} / \tx{multiplier}} {\tx{clock cycle}}
  &= \frac{(\tx{mults/s})/\tx{\#multipliers}}{f}
  \,,
\end{align}
where mults/s above is measured by taking the number of multiplications required
to carry out an execution using conventional algebra and dividing it by the
measured execution time, \#multipliers is the number of instantiated multipliers
in the design, and $f$ is the clock frequency that the hardware design is
operating at.

The throughput metric in \eq{kmm:in-mu} measures the number of $\win$-bit
multiplications being performed, where $\win$ is the algorithm input bitwidths.
However, in order to execute \kmm in hardware, the algorithm input bitwidths
$\win$ must be larger than the multiplier bitwidths,
and the number of larger $\win$-bit multiplications that can be performed per
multiplier will be lower than the actual effective number of multiplications
being performed per multiplier.
Therefore, the maximum achievable value for the metric from \eq{kmm:in-mu} will
vary depending on the input bitwidths $\win$ and is not ideal for reflecting the
true amount of computational work being performed per multiplier regardless of
the input widths.

To address this, we can instead measure \eq{kmm:in-mu} directly in terms of
effective $\wm$-bit multiplications being performed per multiplier, where $\wm$
may be smaller than the algorithm input bitwidths $\win$.
This derives the following metric for
measuring the true amount of effective multiplications being performed per
multiplier regardless of the algorithm input bitwidths $\win$:
\begin{align}  \label{kmm:mu}
  \frac{\wm\tx{-bit mults} / \tx{multiplier}} {\tx{clock cycle}}
  &= \frac{(\wm\tx{-bit mults/s})/\tx{\#multipliers}}{f}
  \,,
\end{align}
where $\wm$-bit mults/s above is measured by taking the number of $\wm$-bit
multiplications required to carry out an execution on $\ww$-bit inputs using
conventional algebra and dividing it by the measured execution time,
\#multipliers is the number of instantiated multipliers in the design, and $f$
is the clock frequency that the hardware design is operating at.
Conventional algorithms used in prior work to perform larger $\ww$-bit
multiplications on smaller $\wm$-bit multipliers are the \sm or \mmn algorithms
(Algorithm \ref{kmm:alg:smn} and \ref{kmm:alg:mmn}).
The number of $\wm$-bit multiplications required to carry out a larger $\win$-bit multiplication using
conventional algebra (i.e. \sm or \mmn) is
equal to the number of $\win$-bit multiplication in the execution times $4^r$,
where $r$ is equal to:
\begin{align}
r = \ceil{\tx{log}_2n} = \ceil{\tx{log}_2\ceil{\win/\wm}}
    \,.
\end{align}

The limit (also referred to as the roof) of the metric in \eq{kmm:mu} when executing
the conventional \mmn algorithm
in hardware is then the following since it
has no algebraic optimizations for reducing the computational complexity:
\begin{align}  \label{kmm:mmn-mu-roof}
  \alg.\MMN'*?\nn/\ww\spc\frac{\wm\tx{-bit mults} / \tx{multiplier}} {\tx{clock cycle}} \tx{ roof} = 1
    \,.
\end{align}
In contrast, the \kmm algorithm requires only $3^r$ smaller-bitwidth
multiplications to form every $\win$-bit product rather than $4^r$ as in \mmn.
Therefore, the multiplier compute efficiency can reach the following limit in
\fpKmm architectures:
\begin{align} \label{kmm:mu-roof}
  \alg.\KMM'*?\nn/\ww\spc\frac{\wm\tx{-bit mults} / \tx{multiplier}} {\tx{clock
  cycle}} \tx{ roof} = \lbb\frac{4}{3}\rbb^{r}
    \,.
\end{align}

\subsection{Area Unit (\autx) Compute Efficiency}
\label{kmm:sec:au}
In this subsection, we define a performance-per-area metric in \eq{kmm:au-roof}
that accounts for the area overhead of registers, adder units, and multipliers
all in a single unit of comparison based around the area of a full adder.
Using this abstracted method for modelling the circuit area allows for a general
complexity analysis that is less biased towards one specific implementation
platform or technology.

We first derive the relative area of adders and registers by modeling that the
area of a $\ww$-bit adder will be approximately equal to the area of $\ww$ full
adders.
We then approximate the area of a $\ww$-bit flip-flop/register relative to a
$\ww$-bit adder according to approximate transistor counts of full adders versus
D-flip-flops based on several sources.
While there are different specific
implementations for these components, we use the approximate transistor count
trends for the implementations in prior work \citeArea, where a standard CMOS
full adder uses 28 transistors \cite{moradi2009ultra} and a 1-bit
flip-flop consumes 18-21 transistors \cite{kawai2014fully},
\cite{cai2019ultra} (which we then approximate as 19.5), to arrive at the
general
area estimation shown in \eq{kmm:eq:add-au} and \eq{kmm:eq:ff-au}
of 1 flip-flop equalling the area of approximately $19.5/28 = 0.7$ full adders.
So long as these
area ratios vary within reasonable bounds as found in prior work \citeArea, the
conclusions from our results do not change.

We then model the approximate area of a $\ww$-bit multiplier circuit based on
the area of a $\ww$-bit adder.
While there are different possible multiplier circuit implementations, the area
of multiplier circuits used in practice commonly scale quadratically with the
area of a full adder \multComplexity, \cite{parhami2010computer}.
Furthermore, the \kmmArch architectures are not tied to being implemented using
one specific multiplier circuit type.
Therefore, in order to provide a more general analysis and insight catering to a
broader range of possible \kmmArch implementations, we approximate the area of a
multiplier based on the general trend of equalling the square of the input
bitwidths times the area of a full adder as shown in \eq{kmm:eq:mult-au}. We then
arrive at the following general area approximations:
\begin{subequations} \label{kmm:eq:op-au} \begin{align}
    \label{kmm:eq:add-au} \areaInt{\add{\ww}} &= \ww \au \\
    \label{kmm:eq:ff-au} \areaB{\eqff} &= 0.7\spc \ww \au \\
    \label{kmm:eq:mult-au} \areaInt{\mult{\ww}}    &= \ww^2 \au
   \,.
\end{align} \end{subequations}

Based on this, we can then derive the AU of each architecture by substituting in
the areas from \eq{kmm:eq:op-au} for each of the corresponding hardware components
in the architectures. The area of a baseline \fpMm MXU is then as follows:
\begin{align}
 \label{kmm:eq:mm1-au}
  \areaInt{\alg.\MM'*?\one} = XY\spc\tx{\Area}(\mult{\win} + 3\spc\eqff\bits{\win} 
  \notag\\  +\spc\accum{2\win} )
  \,.
\end{align}
Here, the area of an accumulator is based on Algorithm \ref{kmm:alg:mm1} and its
implementation in \fig{kmm:pes}, where
the number of accumulator registers and $({2\win}{+}{\wa})$-bit accumulation
adders in the MXU are reduced by a factor of $p$.
Based on this, by substituting in the areas in \eq{kmm:eq:op-au} for the adders and
registers forming the accumulators in \fig{kmm:pes}, every $p$ accumulators on
average then contain the following area:
\begin{align} \label{kmm:au-accum}
  p\spc\areaInt{\accum{2\win}} =&\spc (p-1)\spc\areaInt{\add{2\win{+}\wPT}}
  \notag\\ &+\areaInt{\add{2\win{+}\waT} +\ff{2\win{+}\wa}}
  \,.
\end{align}
In \eqs (\ref{kmm:eq:mm1-au}) - (\ref{kmm:au-accum}), $X$ and $Y$ are the MXU
width and height in number of multipliers,
$\wP = \ceil{\tx{log}_2p}$, and $\wa$ is the following additional bitwidth added to
account for accumulation:
\begin{align}
    \wa &= \ceil{\tx{log}_2 X}
    \,.
\end{align}
As discussed in \secn{kmm:sec:system}, the register requirements in
\eq{kmm:eq:mm1-au} are derived from the fact that each PE in the \fpMm MXU will
contain registers for buffering the $a$ and $b$ inputs being multiplied, as well
as one additional $b$ buffer for loading the next $b$ tile into the MXU as the
current tile is being multiplied.

The area of the \fpKsmm architecture, which is a baseline \fpMm MXU using \ksm
multipliers rather then conventional multipliers, is then:
\begin{align}
 \label{kmm:eq:ksMM-au}
    \areaInt{\alg.\ksMM'*?\nn/\win} = XY\spc \tx{\Area}(\alg.\KSM'*?\nn/\win +
    3\spc\eqff\bits{\win}
    \notag\\ + \accum{2\win})
    \,,
\end{align}
where:
\begin{subequations} \label{kmm:eq:ksm-au} \begin{align}
  \area{\alg.\KSM'*?\nn/\win} =\spc & \areaInt{\add{2\win}}
  \notag\\+& 2\spc\areaInt{\add{2\Hw{\win}{+}4} + \add{\Hw{\win}}}
  \notag\\+&\areaInt{\alg.\KSM'*?{n/2}/{\Hwf{\win}}
  + \alg.\KSM'*?{n/2}/{\Hw{\win}{+}{1}}}
    \notag\\+& \areaInt{\alg.\KSM'*?{n/2}/{\Hw{\win}}}
    \\\areaInt{\alg.\KSM'*?1/\win} &= \areaInt{\mult{\win}}
  \,.
\end{align} \end{subequations}
The addition of $c_0$ on line \ref{kmm:line:ksm-add-c2} of Algorithm
\ref{kmm:alg:ksm} is not included in this area estimate because it can be
performed before line
\ref{kmm:line:ksm-add-c1} where $c_0$ will be on $\win$ bits and will not overlap
with $c_1 \ll \win$. Therefore, this addition can be performed at no cost in
hardware by simply concatenating the two terms together.

The area of the \fpKmm architecture is then:
\begin{subequations} \label{kmm:eq:kmm-au} \begin{align}
  \area{\alg.\KMM'*?\nn} =&\spc 2 X \areaInt{\add{\Hw{\win}}}
  \notag\\ +&2 Y \areaInt{\add{2\Hw{\win}{+}{4}{+}\wa}
  + \add{2\win{+}\wa}}
    \notag\\ +&\areaInt{\alg.\KMM'*?{n/2}/{\Hwf{\win}}
     + \alg.\KMM'*?{n/2}/{\Hw{\win}{+}{1}}}
    \notag\\+& \areaInt{\alg.\KMM'*?{n/2}/{\Hw{\win}}}
    \\\areaInt{\alg.\KMM'*?1/\win} &= \areaInt{\alg.\MM'*?\one/\win}
  \,.
\end{align} \end{subequations}
Due to the nature of right/left shifting by a constant offset in custom
hardware, the shift operations in the \fpKsmm and \fpKmm algorithms do not add
additional area in the corresponding architectures.

We can now compare the AU compute efficiency limits of the \fpMm, \fpKsmm, and
\fpKmm architectures using:
\begin{align}
\label{kmm:au-roof}
    \frac{\tx{throughput}/\tx{Area Unit}} {\tx{clock cycle}} \tx{ \ roof}
    &= \frac{\tx{throughput roof}/\areaT{ARCH}}{f}
    \,,
\end{align}
where ARCH represents one of the mentioned architectures.
Throughput roofs are equal for fixed-precision \fpMm, \fpKsmm, and \fpKmm
architectures with equal $X$/$Y$ MXU dimensions. Therefore, the value of
\eq{kmm:au-roof} for each architecture relative to the \fpMm architecture can be
found through the inverse of its AU from \eq{kmm:eq:mm1-au},
\eq{kmm:eq:ksMM-au}, or \eq{kmm:eq:kmm-au} relative to the inverse of the \fpMm
AU in \eq{kmm:eq:mm1-au} as plotted later in \Figure \ref{kmm:fig:au}.

\section{Results}
\label{kmm:sec:results}

\subsection{Evaluation Metrics}
\label{kmm:metrics}
In \secn{kmm:sec:results}, we compare the \kmmArch architectures against other
designs using the multiplier and Area Unit compute efficiency metrics defined in
\eqs (\ref{kmm:mu}) and (\ref{kmm:au-roof}) from \sections \ref{kmm:sec:mu} and
\ref{kmm:sec:au}, respectively. These are both used to compare an architecture's
throughput per area capabilities regardless of the clock frequency.

Additionally, the multiplier compute efficiency also measures the amount of
computational work being performed per compute area regardless of the clock
frequency \textit{or input bitwidths}.
This is an important quality because prior works using the same compute platform
as us for evaluation only evaluate throughput for input bitwidths $\win$ that
are equal to the multiplier bitwidths $\wm$.
However, in order to execute \kmm in hardware, the input bitwidths $\win$ must
be larger than the multiplier bitwidths.
Therefore, to fairly compare the performance of the prior works against our
\psKmm architecture, we need to use a performance metric with a maximum
achievable value that does not change regardless of the input bitwidths $\win$
being executed, which is not the case for the GOPS metric.
Furthermore, the multiplier compute efficiency is also useful for comparison
with prior works because it is measurable using only throughput, number of
multipliers, and frequency, which are commonly provided or derivable in prior
works.

The Area Unit compute efficiency metric also accounts for the area overhead of
registers and adder units and provides a more general abstracted method for
modelling the circuit area that is less biased towards one specific
implementation platform or technology. However, it is only useful for
comparing
architectures which compute on inputs of the same bitwidth, and it is only
derivable when knowing not only the number of multipliers used in an
architecture, but also the number of adders and registers which is information
that is not readily available from prior works, but we can use
it to model the efficiencies of the fixed-precision \fpKmm architecture against
our baseline designs which we know all of these details about.

\label{kmm:results}
\label{kmm:sec:tables}

\subsection{Comparison to Prior Work}
Although the theoretical concepts presented in this work are general and
applicable to both custom integrated circuits and FPGA implementations, our
example \psKmm implementations were validated on FPGA, and we therefore
compare against state-of-the-art prior works that are also evaluated on FPGA.

As discussed in \secn{kmm:sec:system},
we use an ML accelerator system design based on the one from our previous work
\cite{pogue2024fast}, which has open-source code available \cite{ffip-source},
to house and evaluate our example \psKmm and baseline MXU architectures.
Full system-level validation of the experimental accelerator as integrated into
the system from our previous work \cite{pogue2024fast} has been done on an Arria
10 SoC Developement Kit \cite{sx-dev-kit} containing the \sx device by measuring
throughput in real-time. However, this device contains fewer soft logic
resources than the \gx used in the prior works we compare against, and we
generate compilation results for our design on the same \gx device used in prior
works for a more fair and consistent comparison. Throughput values of our
designs on the \gx device are then calculated using an accurate throughput
estimation model based on our highly deterministic and time-predictable system
implementation, which accurately predicts actual throughputs measured on the \sx
device available to us.
Tables \ref{kmm:tab:first}-\ref{kmm:tab:last} show throughputs for ResNet
\cite{kaiming2016deep} neural network models.



\begin{table*}[]\centering
\sisetup{round-precision=2}          

\caption{Proposed precision-scalable \psKmm and baseline \psMm systolic-array architectures integrated into a deep learning accelerator system compared with each other and prior state-of-the-art deep learning accelerators on \gx FPGA.}
\label{kmm:tab:first}
\label{kmm:tab:kmm-MM}
\label{kmm:tab:8}
\scriptsize
\begin{threeparttable}
  \begin{tabular}{|>{\raggedleft}p{2.5cm}|AA|DA|DC||VVV|HHHVVV|}\toprule
    \arrayrulecolor{black}
    &\multicolumn{2}{c|}{TNNLS '22 \cite{liu2022toward}} &\multicolumn{2}{c|}{TCAD '22 \cite{fan2022fpga}} &\multicolumn{2}{c||}{Entropy '22 \cite{an2022opencl}} &\multicolumn{3}{c|}{\alg.\MM/{\win,8} 64\by64} &\multicolumn{3}{H}{\alg/{\win,8} 64\by64} &\multicolumn{3}{c|}{\alg/{\win,8} 64\by64}                                                      \\
\toprule
\arrayrulecolor{black}
DSP optimization \tnote{1}               &\multicolumn{2}{c|}{Yes}    &\multicolumn{2}{c|}{Yes}        &\multicolumn{2}{c||}{No}                                              &\multicolumn{3}{c|}{Yes}       &\multicolumn{3}{H}{No}           &\multicolumn{3}{c|}{Yes}       \\
\arrayrulecolor{black!30}\midrule
\arrayrulecolor{black}
DSPs                               &\multicolumn{2}{c|}{1473} &\multicolumn{2}{c|}{1473}     &\multicolumn{2}{c||}{1503}                                          &\multicolumn{3}{c|}{1056}    &\multicolumn{3}{H}{1518}        &\multicolumn{3}{c|}{1056}    \\
\arrayrulecolor{black!30}\midrule
\arrayrulecolor{black}
ALMs                               &\multicolumn{2}{c|}{304K} &\multicolumn{2}{c|}{304K}     &\multicolumn{2}{c||}{303K}                                          &\multicolumn{3}{c|}{243K}    &\multicolumn{3}{H}{153K}        &\multicolumn{3}{c|}{250K}    \\
\arrayrulecolor{black!30}\midrule
\arrayrulecolor{black}
Registers                          &\multicolumn{2}{c|}{889K} &\multicolumn{2}{c|}{890K}     &\multicolumn{2}{c||}{-}                                             &\multicolumn{3}{c|}{556K}    &\multicolumn{3}{H}{390K}        &\multicolumn{3}{c|}{562K}    \\
\arrayrulecolor{black!30}\midrule
\arrayrulecolor{black}
Memories                           &\multicolumn{2}{c|}{2334} &\multicolumn{2}{c|}{2334}     &\multicolumn{2}{c||}{1953}                                          &\multicolumn{3}{c|}{2713}    &\multicolumn{3}{H}{2446}        &\multicolumn{3}{c|}{2713}    \\
\arrayrulecolor{black!30}\midrule
\arrayrulecolor{black}
Frequency (MHz)                    &\multicolumn{2}{c|}{200}  &\multicolumn{2}{c|}{220}      &\multicolumn{2}{c||}{172}                                           &\multicolumn{3}{c|}{320}     &\multicolumn{3}{H}{335}         &\multicolumn{3}{c|}{326}     \\
\arrayrulecolor{black!30}\midrule
\arrayrulecolor{black}
Model                              &ResNet-50      &VGG16     &Bayes ResNet-18  &Bayes VGG11 &R-CNN (ResNet-50) &R-CNN (VGG16)  &ResNet-50 &ResNet-101 &ResNet-152   &ResNet-50 &ResNet-101 &ResNet-152      &ResNet-50 &ResNet-101 &ResNet-152      \\
\arrayrulecolor{black!30}\midrule
\arrayrulecolor{black}
Input bitwidth ($\win$)               &8              &8         &8                &8           &8                 &8      &1-8 / 9-16  &1-8 / 9-16 &1-8 / 9-16     &1-8 / 9-14 / 15-16  &1-8 / 9-14 / 15-16  &1-8 / 9-14 / 15-16                                  &1-8 / 9-14 / 15-16  &1-8 / 9-14 / 15-16  &1-8 / 9-14 / 15-16                                       \\
\arrayrulecolor{black}\midrule
\tabThpt                              &1519           &1295      &1590             &534         &719               &865    &2108 / 527  &2304 / 576  &2390 / 598     &2200 / 735 / 552 &2412 / 804 / 603  &2502 / 834 / 626            &2147 / 716 / 537 &2347 / 782 / 587  &2435 / 812 / 609                                              \\
\arrayrulecolor{black!30}\midrule
\arrayrulecolor{black}
\arrayrulecolor{black}
\kmmMacUt                             &0.645          &0.550     &0.639            &0.206       &0.696             &0.837  &0.792 / 0.792  &0.865 / 0.865  &0.898 / 0.898        &0.792 / \textbf{1.055} / 0.792 &0.865 / \textbf{1.154} / 0.865 &0.898 / \textbf{1.197} / 0.898&0.792 / \textbf{1.055} / 0.792 &0.865 / \textbf{1.154} / 0.865 &0.898 / \textbf{1.197} / 0.898                     \\
\arrayrulecolor{black}
\bottomrule
\end{tabular}
\begin{tablenotes}
\DSPnote
\kmmMacUtExpl
\end{tablenotes}
\end{threeparttable}
\end{table*}


\newcolumntype{E}{>{\centering\arraybackslash}p{1.2cm}}

\begin{table*}[]\centering

  \caption{Comparison of an FFIP \cite{pogue2024fast} systolic array, which
  doubles performance per MAC unit, with combined FFIP+\psKmm precision-scalable
  systolic arrays when integrated into deep learning accelerator systems on \gx
  FPGA.}
\label{kmm:tab:ffip}
\scriptsize
\begin{threeparttable}
  \begin{tabular}{|>{\raggedleft}p{2.5cm}|VVV|VVV|VVV|}\toprule
\arrayrulecolor{black}
                                   &\multicolumn{3}{c|}{TC '24 \cite{pogue2024fast} (FFIP 64\by64)}    &\multicolumn{3}{c|}{\alg.\kffip/{\win,8} 64\by64}                                                   &\multicolumn{3}{c|}{\alg.\kffip/{\win,8} 64\by64}                                                    \\
\toprule
 DSP optimization \tnote{1}                &\multicolumn{3}{c|}{No}                                             &\multicolumn{3}{c|}{No}                                                                             &\multicolumn{3}{c|}{Yes}                                                                              \\
\arrayrulecolor{black!30}\midrule
\arrayrulecolor{black}
DSPs                               &\multicolumn{3}{c|}{\eDSPs}                                        &\multicolumn{3}{c|}{1072}                                                                          &\multicolumn{3}{c|}{552}                                                                           \\
\arrayrulecolor{black!30}\midrule
\arrayrulecolor{black}
ALMs                               &\multicolumn{3}{c|}{\eALMs}                                        &\multicolumn{3}{c|}{133K}                                                                          &\multicolumn{3}{c|}{205K}                                                                           \\
\arrayrulecolor{black!30}\midrule
\arrayrulecolor{black}
Registers                          &\multicolumn{3}{c|}{\eRegs}                                        &\multicolumn{3}{c|}{334K}                                                                          &\multicolumn{3}{c|}{502K}                                                                           \\
\arrayrulecolor{black!30}\midrule
\arrayrulecolor{black}
Memories                           &\multicolumn{3}{c|}{\eMems}                                        &\multicolumn{3}{c|}{2445}                                                                          &\multicolumn{3}{c|}{2713}                                                                           \\
\arrayrulecolor{black!30}\midrule
\arrayrulecolor{black}
Frequency (MHz)                    &\multicolumn{3}{c|}{\eFreq}                                        &\multicolumn{3}{c|}{353}                                                                           &\multicolumn{3}{c|}{341}                                                                            \\
\arrayrulecolor{black!30}\midrule
\arrayrulecolor{black}
Model                              &ResNet-50            &ResNet-101          &ResNet-152              &ResNet-50 &ResNet-101 &ResNet-152                                                                  &ResNet-50 &ResNet-101 &ResNet-152                                                                   \\
\arrayrulecolor{black!30}\midrule
\arrayrulecolor{black}
Input bitwidth ($\win$)               &8&8&8                                                           &1-8 / 9-14 / 15-16  &  1-8 / 9-14 / 15-16  &  1-8 / 9-14 / 15-16                                   &1-8 / 9-14 / 15-16  &  1-8 / 9-14 / 15-16  &  1-8 / 9-14 / 15-16                                    \\
\arrayrulecolor{black}\midrule
\tabThpt                              &\eResNetAGOPS        &\eResNetBGOPS       &\eResNetCGOPS        &2325 / 775 / 581 &2542 / 847 / 635  &2637 / 879 / 659                                              &2246 / 749 / 562 &2455 / 818 / 614  &2547 / 849 / 637                                               \\
\arrayrulecolor{black!30}\midrule
\kmmMacUtFFIP                             &1.521  &1.655 &1.707                                           &1.536 / \textbf{2.048} / 1.536   &1.679 / \textbf{2.239} / 1.679   &1.742 / \textbf{2.322} / 1.742 &1.536 / \textbf{2.048} / 1.536   &1.679 / \textbf{2.239} / 1.679   &1.742 / \textbf{2.322} / 1.742  \\
\arrayrulecolor{black}
\bottomrule
\end{tabular}
  \begin{tablenotes}
\DSPnote
\kmmMacUtExpl
\end{tablenotes}
\end{threeparttable}
\end{table*}

In Table \ref{kmm:tab:8}, the number of multipliers in the work from An \ea
\cite{an2022opencl} is calculated as $\#DSPs \times 2$, where each DSP in the
Intel/Altera FPGAs contains two 18-bit multipliers \cite{intel-dsp}.
The works from Liu \ea \cite{liu2022toward} and Fan \ea \cite{fan2022fpga} in
Table \ref{kmm:tab:8} implement a similar method as in the work from Langhammer
\ea \cite{langhammer2019extracting} to
pack two 8-bit multiplications onto each 18-bit
multiplier in the DSPs at the cost of additional ALMs and registers, and
therefore $\#multipliers = \#DSP
\times 4$ in those works. Our architectures in Table \ref{kmm:tab:8} contains 64\by64
+ 64 multipliers, where 64\by64 multipliers are used in the MXU, while the
remaining 64 are located outside the MXU in the Post-GEMM Unit
\cite{pogue2024fast} for performing inter-layer quantization rescaling
functions.
This is also how the number of multipliers is calculated in the architectures in
Table \ref{kmm:tab:ffip}, except there the MXUs contain 64\by32+32
multipliers due to using the FFIP method \cite{pogue2024fast}.
For the multipliers located in the MXU of our designs in Tables
\ref{kmm:tab:8}-\ref{kmm:tab:ffip}, we also implement
a similar method as in the work from Langhammer
\ea \cite{langhammer2019extracting} to
pack two smaller-bit multiplications onto each 18-bit
multiplier in the DSPs.
However, we leave one FFIP+KMM design in Table
\ref{kmm:tab:ffip} without this optimization for a more fair comparison to the
FFIP design in our prior work \cite{pogue2024fast} that did not
implement this optimization.

Table \ref{kmm:tab:8} compares the \psKmm architecture with state-of-the-art
accelerators evaluated on the same FPGA family for the same instantiated
multiplier bitwidths and similar neural network models. The proposed \psKmm
architecture is very efficient,
achieving the highest throughput
and
compute efficiency compared to the prior works in Table \ref{kmm:tab:8}.
The \psKmm design here achieves compute efficiencies
approaching the \psKmm?2 limit of 1.33 when
executing on bitwidths in the range of 9-14 bits that is derived in
\eq{kmm:mu-roof} and
surpasses the limit of 1 in prior works
that is derived in \eq{kmm:mmn-mu-roof}.

It is also noted that the proposed systolic arrays in Tables \ref{kmm:tab:8} and
\ref{kmm:tab:ffip} that are integrated into a full accelerator
system include
a number of other components such as memory subsystems and control as
described in our prior work \cite{pogue2024fast}, and these
other system components form the frequency-limiting critical path
as opposed to the proposed systolic-array architectures.

Table \ref{kmm:tab:ffip} shows an example of how \kmm can be combined with other
algebraic techniques to further increase compute efficiency limits.
FFIP \cite{pogue2024fast} provides a way to reduce the number of required
multiplications by a factor of 2, by trading half the multiplications for cheap
low-bitwidth additions.
Because the number of required multiplications is reduced by 2, the limit for
the multiplier compute efficiency metric in \eq{kmm:mu-roof} becomes 2 for FFIP,
and $(8/3)^r$ for FFIP+\kmm.
In Table \ref{kmm:tab:ffip}, we combine \psKmm with FFIP \cite{pogue2024fast} by
using an FFIP MXU as the base MXU in the \psKmm architecture instead of a
conventional \fpMm MXU to further increase the compute efficiency compared to
standalone FFIP.
The FFIP+\kmm architectures in Table \ref{kmm:tab:ffip} have additional memory
resources instantiated compared to the FFIP-only design in order to support
inference on up to 16-bit inputs, and this also adds a penalty in the soft
logic resources and clock frequency.
However, the multiplier compute efficiency of the FFIP+\kmm designs
surpass the FFIP limit of 2, and approach the FFIP+\kmm?2 limit of 2.67.

\subsection{Comparison to Baseline Designs}
\subsubsection{Precision-Scalable Architectures}

\newcolumntype{X}{>{\centering\arraybackslash}p{.57cm}}
\newcolumntype{Y}{>{\centering\arraybackslash}p{.95cm}}

\begin{table*}[]\centering
\sisetup{round-precision=2}          

\caption{Comparison of proposed fixed-precision \fpKmm and baseline \fpMm and \fpKsmm systolic-array architectures in isolation (without integration into a deep learning accelerator system) on Agilex 7 FPGA.}
\label{kmm:tab:fp}
\label{kmm:tab:last}
\scriptsize
\begin{threeparttable}
  \begin{tabular}{|>{\raggedleft}p{2.5cm}|XX|YY|Y||XX|YY|YHH|}\toprule
    \arrayrulecolor{black}
    &\alg.\MM?1/{32} 32\by32 &\alg.\MM?1/{32} 32\by32  &\alg.\ksMM?2/{32} 32\by32 &\alg.\ksMM?2/{32} 32\by32  &\alg.\KMM?2/{32} 32\by32    &\alg.\MM?1/{64} 32\by32 &\alg.\MM?1/{64} 32\by32  &\alg.\ksMM?4/{64} 32\by32 &\alg.\ksMM?4/{64} 32\by32  &\alg.\KMM?4/{64} 32\by32  &\alg.\MM/{\win,8}?2 64\by64 &\alg.\KMM/{\win,8}?2 64\by64
    \\
\toprule
Input bitwidth                     &32 &32  &32 &32  &32    &64 &64  &64 &64  &64           &1-16 / 17-30 / 31-32  &1-16 / 17-30 / 31-32 \\
\arrayrulecolor{black!30}\midrule
\arrayrulecolor{black}
DSPs                               &2048 &2048  &1536 &1536  &1536    &8704 &8704  &4608 &4608 &4608    &512 &512  \\
\arrayrulecolor{black!30}\midrule
\arrayrulecolor{black}
ALMs                               &64K &69K  &138K &147K &68K   &240K &266K  &554K &557K  &212K    &14 &15  \\
\arrayrulecolor{black!30}\midrule
\arrayrulecolor{black}
Registers                          &165K &225K&  306K &481K  &257K    &237K &712K  &447K &1126K  &806K   &50K &51K \\
\arrayrulecolor{black!30}\midrule
\arrayrulecolor{black}
Frequency (MHz)                    &450 &569  &386 &537  &622   &203 &341  &147 &345  &552    &708 &716  \\
\arrayrulecolor{black}\midrule
\arrayrulecolor{black}
Throughput roof (GOPS)             &922 &1165  &791 &1100  &1274    &416 &698  &302 &707  &1131        &2200 / 735 / 552 &2412 / 804 / 603 \\
\arrayrulecolor{black}
\bottomrule
\end{tabular}
\begin{tablenotes}
\item All designs in this table consume 0 memory resources and are synthesized for an Agilex 7 AGIA040R39A1E1V device.
\end{tablenotes}
\end{threeparttable}
\end{table*}

Table \ref{kmm:tab:kmm-MM} includes the resource usage and performance comparison
between the proposed \psKmm and the baseline \psMm architectures. The multiplier
compute efficiency of \psKmm surpasses that of the baseline \psMm architecture when
executing on bitwidths in the range of 9-14 bits, achieving compute efficiencies
approaching the \psKmm?2 limit of 1.33 that is derived in \eq{kmm:mu-roof} and
surpassing the limit of 1 of the baseline \psMm architecture and prior works
that is derived in \eq{kmm:mmn-mu-roof}, validating \kmm's ability to increase
compute efficiency as expected from our analysis. This is also reflected in the
GOPS from Table \ref{kmm:tab:kmm-MM}, where the \psKmm architecture achieves a
1.33\x speedup over \psMm for input bitwidths in the range of 9-14 bits.

\begin{figure}[]
  \vspace*{-0.2cm}
  \centering
  \includegraphics[scale=0.0525]{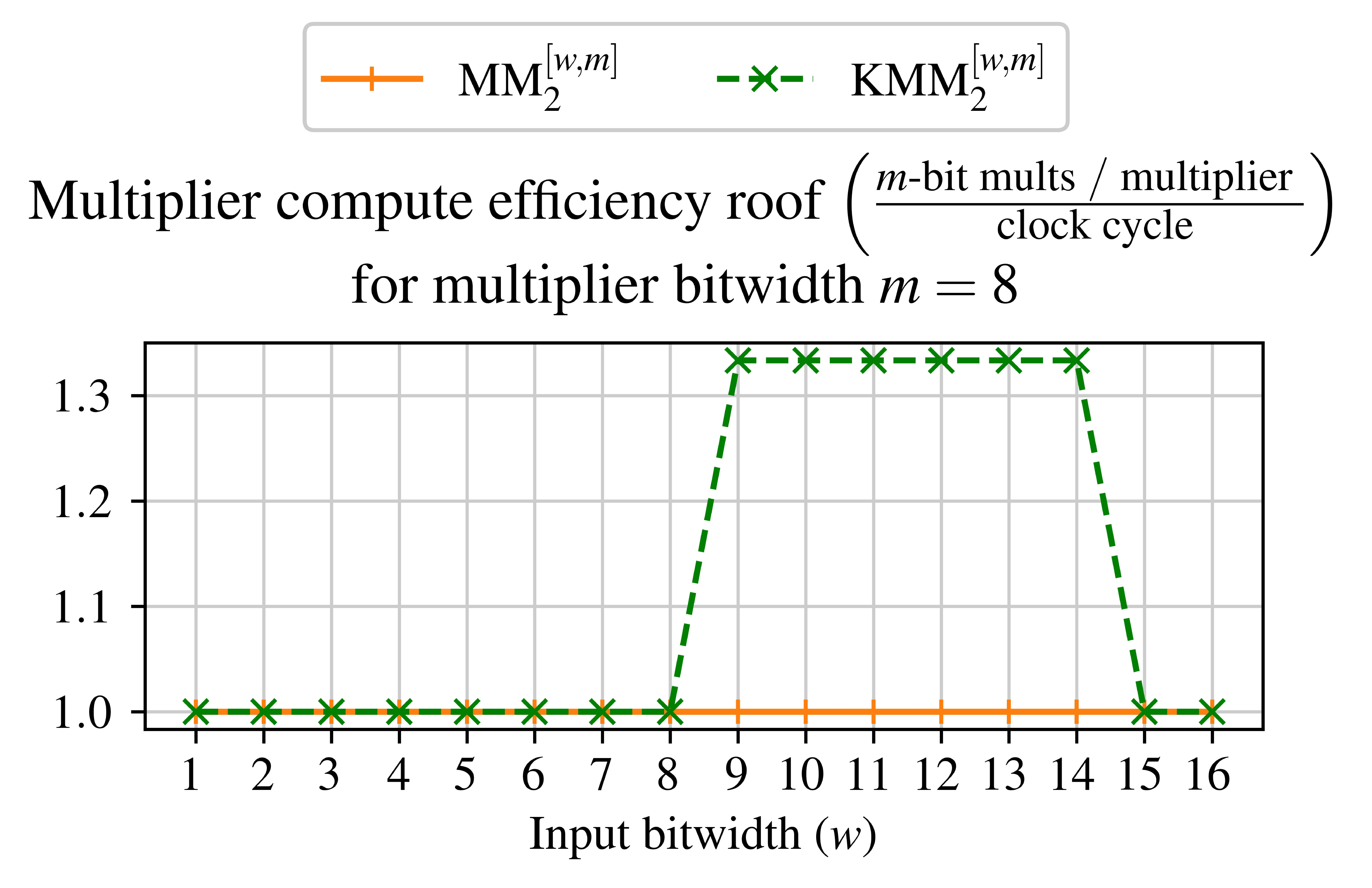}
  \vspace*{-0.7cm}
  \caption{Maximum achievable multiplier compute efficiencies (derived in
    \secn{kmm:sec:mu}) for the precision-scalable \psMm?2 and \psKmm?2
    architectures.}
  \label{kmm:fig:mu}
\end{figure}

For illustration, \fig{kmm:fig:mu} plots the limits of the multiplier compute
efficiency metric defined in \eq{kmm:mu} from \secn{kmm:sec:mu} for the
precision-scalable \psKmm?2 architecture compared to the conventional
precision-scalable \psMm?2 architecture for $X = Y = 64$. As shown, the \psKmm
architecture surpasses the \psMm architecture's limit of 1 for this metric,
extending the limit to 1.33 for bitwidths 9-14 since the \psKmm?2 algorithm
requires only 3 $\wm$-bit multiplications for every $\win$-bit product rather
than 4 as in the \psMm?2 algorithm.

\subsubsection{Fixed-Precision Architectures}
Table \ref{kmm:tab:fp} shows synthesis results on a modern Agilex 7 FPGA device for
baseline \mm, \ksmm, and proposed \kmm systolic-array architectures in isolation
(not
integrated into a deep learning accelerator) for different input bitwidths and
levels of \ksm and \kmm recursion. The input bitwidths are intentionally larger
than the DSP units' native multiplier bitwidths and are
chosen to allow for larger multiplications
to be broken down into smaller multiplications of bitwidths at or just below the
native widths supported by the DSPs, which house 18-bit multipliers.
It is expected that the larger-bit multiplications in the \fpMm designs
will be mapped to smaller 16-bit multipliers, and onto fewer 16 to 18-bit
multipliers in the \fpKmm and \fpKsmm designs.

The reduction in multiplication complexity of \kmm and \ksmm achieved through breaking down
larger multiplications into smaller-bitwidth multiplications can be seen relative to
conventional approaches (evaluated through the \mm architectures)
by comparing the reduction in number of DSP units for the \kmm and \ksmm designs
relative to \mm. Furthermore, the reduction in addition complexity of \kmm
relative to \ksmm can be seen in the reduction in ALMs in the \fpKmm
architectures compared to the \fpKsmm architectures.

The \mm and \ksmm architectures innately have a lower clock
frequency than \kmm because it is expected that each multiplication being
performed in the PEs require $n^2$ or $n\klog$ DSP units, respectively, whereas
the \kmm designs require only 1 DSP unit in each individual \kmm systolic-array
PE.
This leads to a less localized design.
In contrast, the \kmm design uses multiple independent systolic arrays
requiring 1 DSP unit per multiplication to perform a single 16 to 18-bit
multiplication, and the DSPs in each systolic array do not require
interconnections with the DSPs in other systolic arrays, leading to a more
localized design.
Due to this, we provide results of two design variants for each of the \mm and
\ksmm architectures, where one variant contains additional pipelining
registers added into the PE datapaths such that the clock frequency can reach
closer to that of the \kmm designs. However, it can be seen that the \mm and
\ksmm designs are still unable to match the frequency of \kmm even with
extra pipelining registers, especially for the 64-bit input designs.

In summary, the trend in Table \ref{kmm:tab:fp} is that the \kmm designs may
contain more register resources than the \mm and \ksmm designs depending on the
amount of pipelining registers used, however, the \kmm designs use significantly
fewer ALM resource than the \ksmm designs, significantly fewer DSP units than
the \mm designs, and achieve significantly higher clock frequencies than
both \ksmm and \mm.

\begin{figure}[]
  \vspace*{-0.1cm}
  \centering
  \includegraphics[scale=0.0525]{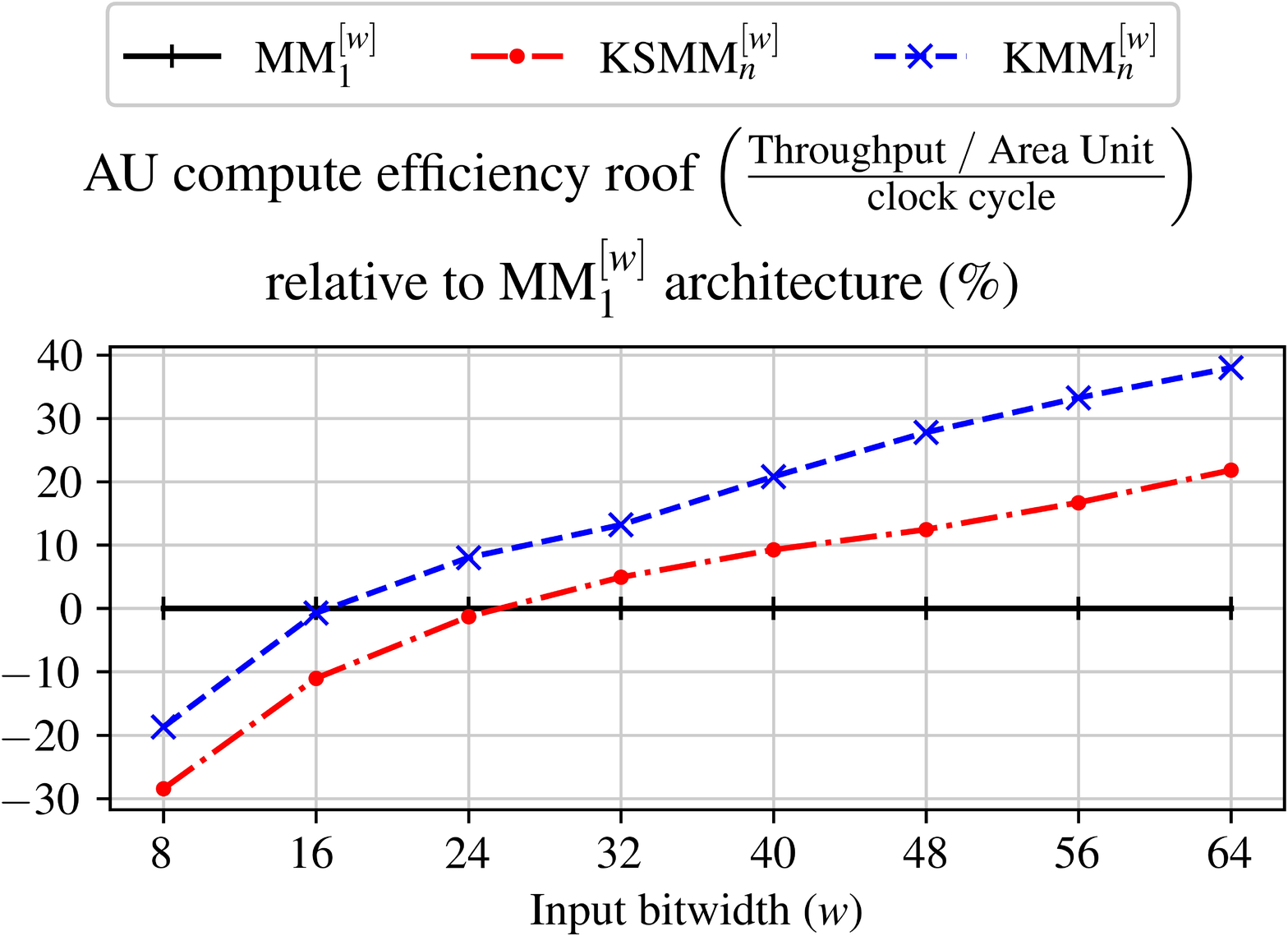}
  \vspace*{-0.6cm}
  \caption{Maximum achievable AU compute efficiencies (derived in
    \secn{kmm:sec:au}) for the fixed-precision \fpMm, \fpKsmm?n, and \fpKmm?n
    architectures.}
  \label{kmm:fig:au}
\end{figure}

\fig{kmm:fig:au} also provides a more general modelling of the performance-per-area
of the \kmm architectures that is less biased towards one specific
implementation platform or technology by plotting the AU compute efficiency
limits derived in \secn{kmm:sec:au}
that can be achieved for the fixed-precision
\fpMm, \fpKsmm, and \fpKmm architectures for different supported fixed-precision
input widths and instantiated multiplier bitwidths for $X = Y = 64$.
The \fpKmm and \fpKsmm architectures for each bitwidth implement as many levels
of Karatsuba recursion as possible while still reducing the area, with a minimum
of least one level of Karatsuba recursion being implemented (even if the one
level has a larger area than using conventional \mm). This results in one
recursion level being implemented in the \fpKsmm architectures for every bitwidth.
For the \fpKmm architectures, this results in one recursion level
for bitwidths 8-32, two recursion levels for bitwidths 40-56, and
three recursion levels for bitwidth 64.

As can be seen, the \fpKmm architecture achieves a higher throughput per Area
Unit than the conventional \fpMm architecture starting sooner at a lower
bitwidth compared to the \fpKsmm architecture, and it is consistently higher
than the \fpKsmm architecture across all input/multiplier bitwidths.

\section{Conclusion}
In this work, we propose the extension of the scalar Karatsuba multiplication
algorithm to matrix multiplication,
showing how this maintains the reduction in multiplication complexity of the
original Karatsuba algorithm while reducing the complexity of the extra additions.
Furthermore, we propose new matrix multiplication hardware architectures for
efficiently exploiting the proposed algorithm in custom hardware,
showing that they can provide real
area or execution time improvements for integer matrix multiplication compared
to designs implementing scalar Karatsuba or conventional matrix multiplication
algorithms.
The proposed architectures are well suited for increasing the efficiency in
acceleration of modern workloads that can decompose to large matrix
multiplications on integer arithmetic, such as the computationally dominant
portion of convolutional neural networks or the attention mechanism of
transformer models \cite{trans}.
We provide a complexity analysis of the algorithm and architectures
and evaluate the proposed designs both in isolation
and in an end-to-end accelerator system relative to baseline designs and prior
state-of-the-art works, showing how they increase the
performance-per-area of matrix multiplication hardware.

\bibliographystyle{IEEEtran}
\bibliography{IEEEabrv,bibl}

\begin{IEEEbiography}[{\includegraphics[width=1in,height=1.25in,clip,keepaspectratio]{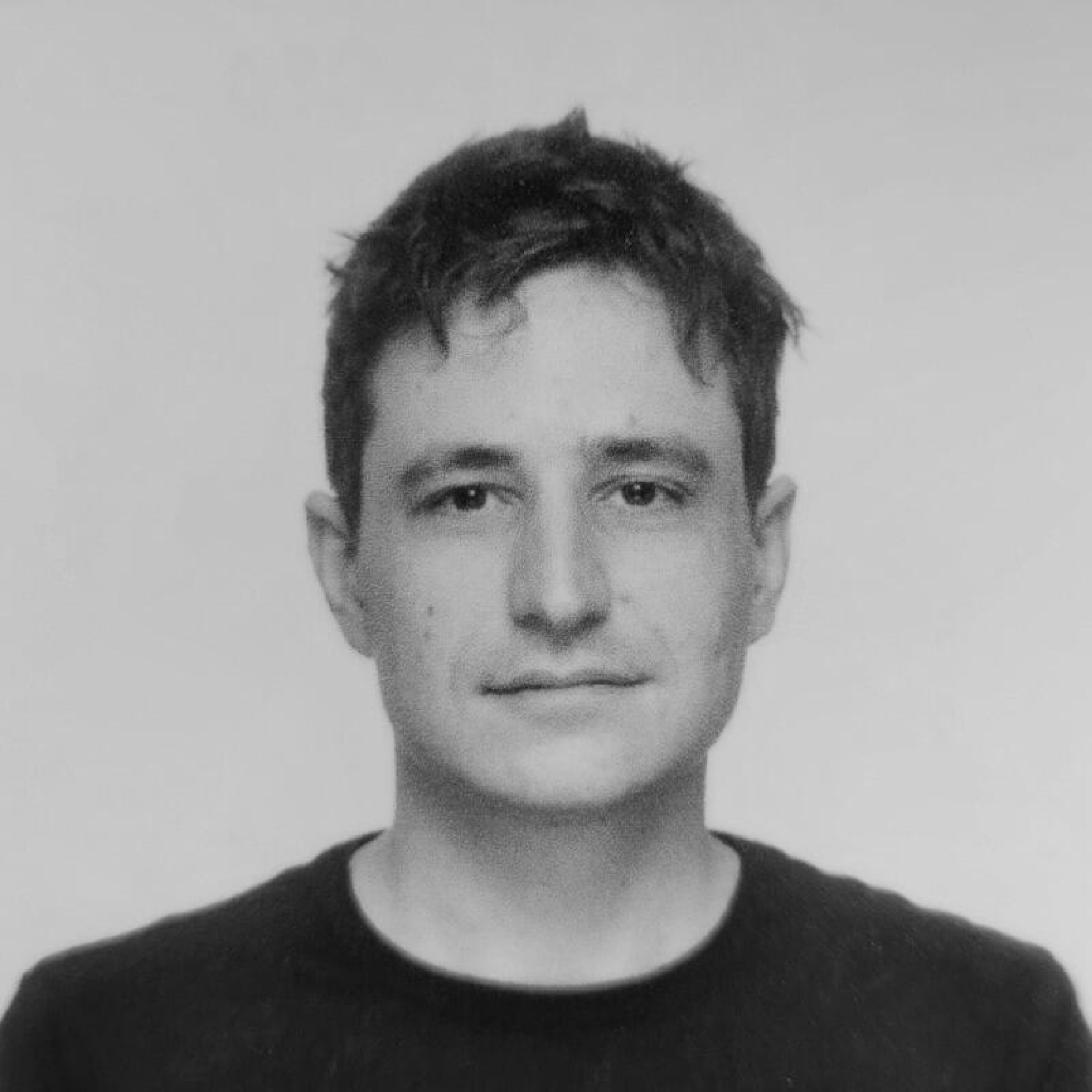}}]{Trevor E. Pogue}
  Trevor E. Pogue received the B.Eng. degree in Electrical Engineering and the M.A.Sc. degree in Electrical and Computer Engineering from McMaster University, Hamilton, Canada, in 2016 and 2019, respectively. He worked as an intern at Synopsys and AMD in 2018 and 2022-2023, respectively. He is currently a Ph.D. Candidate in the Department of Electrical and Computer Engineering at McMaster University, Hamilton, Canada. His research interests are in the area of hardware acceleration.
\end{IEEEbiography}

\begin{IEEEbiography}[{\includegraphics[width=1in,height=1.25in,clip,keepaspectratio]{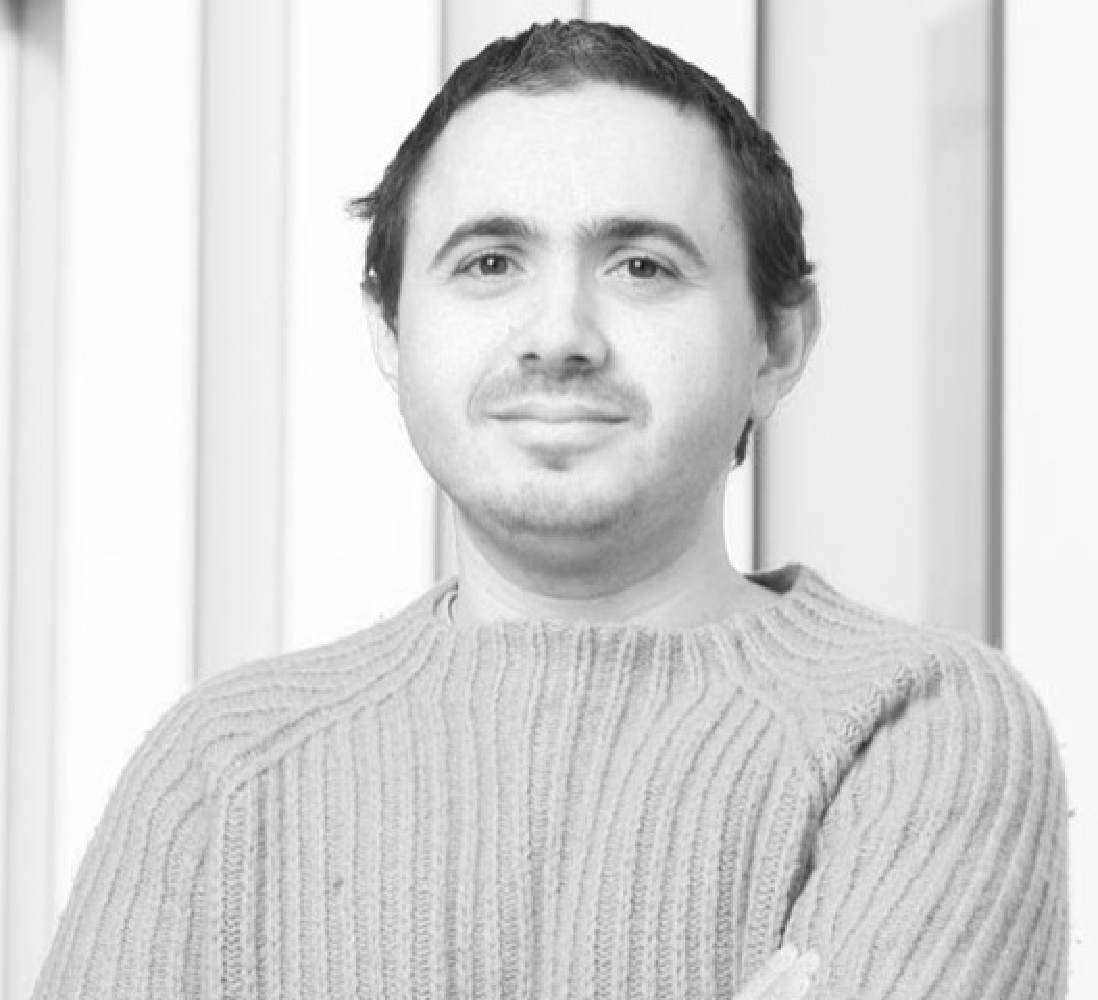}}]{Nicola
    Nicolici}(S’99-M’00-SM'11)
  Nicola Nicolici (S99-M00-SM’11) received the Dipl.Ing. degree in Computer
  Engineering from the “Politehnica” University of Timisoara, Romania, in 1997
  and the Ph.D. degree in Electronics and Computer Science from the University
  of Southampton, U.K., in 2000. He is currently a Professor with the Department
  of Electrical and Computer Engineering, McMaster University, Hamilton, Canada.
  His research interests are in the area of computer-aided design and test.
  He has authored a number of papers in this area.
  Dr. Nicolici was the
  recipient of the IEEE TTTC Beausang Award for the Best Student Paper at the
  International Test Conference in 2000 and the Best Paper Award at the IEEE/ACM
  Design Automation and Test in Europe Conference in 2004.
\end{IEEEbiography}

\end{document}